\RequirePackage{fix-cm}
\documentclass[smallextended]{svjour3}       
\smartqed  
\usepackage{graphicx}
\begin{document}

\title{Study of Energy Extraction and Epicyclic Frequencies in Kerr-MOG~(Modified Gravity) Black Hole
}


\author{Parthapratim Pradhan}
\institute{\at Department of Physics\\
            Hiralal Mazumdar Memorial College For Women \\
           Dakshineswar, Kolkata-700035, India.\\
           \email{pppradhan77@gmail.com}}
\date{Received: date / Revised version: date}

\maketitle

\begin{abstract}
We investigate the energy extraction by the Penrose process in Kerr-MOG black hole~(BH). 
We derive the gain in energy for Kerr-MOG as
\begin{eqnarray}
\Delta {\cal E} \leq \frac{1}{2}\left(\sqrt{\frac{2}{1+\sqrt{\frac{1}{1+\alpha}-\left(\frac{a}{{\cal M}}\right)^2}}
-\frac{\alpha}{1+\alpha} \frac{1}{\left(1+\sqrt{\frac{1}{1+\alpha}-\left(\frac{a}{{\cal M}}\right)^2} \right)^2}}-1\right)
\nonumber
\end{eqnarray}
 Where $a$ is spin parameter, $\alpha$ is MOG parameter and ${\cal M}$ is the Arnowitt-Deser-Misner(ADM) 
mass parameter. When $\alpha=0$, we obtain the gain in energy  for Kerr BH. For extremal Kerr-MOG BH, we determine 
the maximum gain in energy is $\Delta {\cal E} \leq \frac{1}{2} \left(\sqrt{\frac{\alpha+2}{1+\alpha}}-1 \right)$. 
We observe that the MOG parameter has a crucial role in the energy extraction process and it is in fact 
diminishes the value of $\Delta {\cal E}$ in contrast with extremal Kerr BH. Moreover, we derive 
the \emph{Wald inequality and the Bardeen-Press-Teukolsky inequality} for Kerr-MOG BH in contrast with 
Kerr BH. Furthermore, we describe the geodesic motion in terms of three fundamental frequencies: 
the Keplarian angular frequency, the radial epicyclic frequency and the vertical epicyclic frequency. 
These frequencies could be used as a probe of strong gravity near the black holes.
\end{abstract}

\section{Introduction}

Black hole~(BH) is the most facinating as well as compact objects in the universe. It has several 
facinating properties.  Among, one of them is the energy extraction by the Penrose process. 
Classically, it is impossible to extract energy from the non-spinning BH but it is possible 
to extract rotational energy from spinning BH~\cite{jbh,mtw,wald,gero,bk73}. The most important 
feature of rotating/spinning BH is that the presence of the ergosphere while the non-spinning 
BH does not possess such ergo region. Ergosphere is responsible for several important 
phenomena in BH physics.  
The idea of energy extraction was first came to in mind by Roger Penrose in 1969~\cite{rp69,rp71}. He 
first showed how the ergosphere could be in principle exploited to extract the rotational 
energy from the BH.  

Another important feature of spinning BH is that the Killing vector $\xi^{\mu}=\partial_{0}$ which 
is time-like at $\infty$ becomes space-like in the ergosphere~(i.e. the toroidal space between 
the event horizon and the stationarity limit surface on which the components of the axially 
symmetric metric $g_{00}=0$). Moreover, the existence of particle orbits with negative total energy which could 
be measured from infinity. This energy is defined as $E=-p_{\mu}\xi^{\mu}$, where $\xi^{\mu}$ is the 
four momentum of the test particle. Outside the ergosphere (where $\xi^{\mu}$ is time-like) the energy must 
be positive, however inside the ergosphere~(where $\xi^{\mu}$ is time-like) the energy has the nature of a 
spatial component of momentum and have either sign~\cite{jbh,wald,rp71,sch}.

Penrose first proposed that one can take the advantage of these negative orbits to extract rotational energy 
from the BH. The process could be understood shortly as follows.
In this process, a particle falls into the ergosphere from infinity. Then it decays into
two fragments. One fragment escapes to infinity and other fragment plunges through the event horizon 
into the BH. Both the energy and the momentum conserved in this hypothetical process. Therefore, one 
can extract the rotational energy from the BH. It should be noted that in the ergosphere, the Killing 
vector $\partial_{0}$ becomes spacelike as said previously and similarly the conserved component, 
$p_{0}$, of the four-momentum. Therefore when an observer observes the toroidal space from infinity
he/she could be discerned that the energy of the particle becomes negative. Due to this negative energy, one 
could be able to extract both the energy and the angular momentum from the BH. However the area of 
BH's event horizon never decreases. Either it must be increases or remains constant.

The first motivation comes from the work of Penrose who showed how to extract energy from a 
Kerr BH. Here we would like to extend  this work for Kerr-MOG BH. Because this BH is described by 
three parameters i.e. namely the spin parameter $a$, the ADM mass parameter ${\cal M}$ 
and the MOG parameter $(\alpha)$. Whereas the Kerr BH consists of only two parameters 
i.e. the ADM mass parameter and the spin parameter. 
Due to the presence of the deformation parameter what will be the change in the 
``gain in energy expression'' in extraction process in contrast to the Kerr BH. This is the
primary motivation behind this work. We also investigated the Wald inequality which gives 
the energy limits on the energy extraction process. Futhermore, we have discussed the 
Bardeen-Press-Teukolsky inequality. Finally, we have considered the reversible extraction 
of energy and the irreducible mass for Kerr-MOG BH.

What is the problem with Einstein's general theory of relativity~(GTR)? It is an incomplete theory in a
sense that it breaks down at short length scale. It is unnecessary to taken into account the 
quantum  effect. It could not explain the large scale behaviour of gravitational field. 
The lacking of this characteristic features gave birth a new kind of gravity which is 
called MOG. The MOG is formulated by scalar 
field and massive vector field that's why the MOG theory is also called 
the scalar-tensor-vector-gravity~(STVG). The MOG theory correctly interpreted the observations 
of the solar system~\cite{mf}~[See also~\cite{mf1,mf2,mf3,mf5,mf6,mf7,mf8,mf9,mf10}]. 
It also explains the rotation curves of the cluster of galaxies and 
the dynamics of the cluster of galaxies. Moreover, the STVG theory correctly describes the 
power spectrum of matter and the acoustical power spectrum of the cosmic microwave background~(CMB) 
data~\cite{mf}. 

The modified action for the STVG theory is equal to the sum of four actions, namely 
the Einstein-Hilbert action for gravity,  the action for massive vector field,  
the  action for scalar fields and the action for pressure less matter. 
This means that we can derive the equations of motion from an action principle. This theory is also 
covariant and obeys the weak equivalence principle~\cite{mf9}. Like GTR, the MOG theory 
allows to testify the gravitational wave signals~\cite{mf7} and predicts the gravitational 
lensing features of cluster of galaxies. 

There has been compelling evidence of ring down of BH mergers~\cite{mf} and BH shadow~\cite{mf5} 
have been detected in MOG. In some way we have been able to measure the quasi normal mode frequencies 
from a binary BH merger, the shadow produced by massive object and to interpret both of them 
as consistent with the MOG theory. Besides that it must be noted that the above two quantities has 
not been clearly observed till to date~(the QNM of the first GW event is still questionable and the 
first observational results on the BH shadow are coming out in these months)~\cite{eht,bhi}.

The stability properties for MOG has been studied under gravitational perturbation and 
electromagnetic perturbation in Ref.~\cite{mf10}. In this Ref.,  the author also 
calculated the quasi normal modes~(QNM) frequency of static BHs in STVG theory using 
Asymptotic Iteration Method~(AIM). They showed there is a clear distinction 
between MOG QNMs and GR QNMs. They suggested possible experimental detection of QNMs 
frequency using LISA and LIGO data.

The thermodynamic properties of MOG has been explicitly examined in Ref. ~\cite{pp18}.
Where the author studied the outer/inner horizon thermodynamics of MOG and their consequences 
on holographic duality. Entropy product formula of spherically symmetric and axisymmetric MOG 
does depend on the mass parameter hence the product is not a universal quantity. 
The first law is satisfied at the inner horizon and outer horizon for MOG BH. Smarr like formula 
is satisfied for MOG BH. Using Kerr-MOG/CFT~(conformal field theory) correspondence, it was shown that the 
central charges for Kerr-MOG BH is similar to Kerr BH i.e. $c_{L}=12J$. Where $J$ is angular 
momentum.  The dual CFT temperature of Frolov-Thorne thermal vacuum state has been derived for 
extremal Kerr-MOG BH and it was shown that it strictly depends on the MOG parameter. The Cardy 
formula helped us  to derive the microscopic entropy for extremal Kerr-MOG BH 
which was completely in agreement with the macroscopic Bekenstein-Hawking entropy. Therefore 
one may conjectured that in the extremal limit, the Kerr-MOG BH is holographically dual to a 
chiral 2D CFT with central charge $c_{L} = 12J$.

Further motivation for the work comes from the fact that MOG BHs do 
Hawking radiate which is known to be absent for extremal situation because the 
surface gravity (which is computed on the horizon) measures equilibrium 
temperature for the thermal distribution of the radiation. It was proved in~\cite{wi86}
that at a finite advanced time  no continuous process can make a nonextremal BH to extremal 
BH in a finite number of process by lossing its traped surface. Analogously, one cannot make 
a nonextremal Kerr MOG BH to a extremal Kerr-MOG BH in a finite of steps.

Now we must mention here the several important works regarding the MOG theory. 
In \cite{mf}, the basic MOG formulation  i.e. STVG theory was introduced. 
In ~\cite{mf1}, the observational test of galaxy rotation curves in the 
MOG weak field approximation was discussed. In ~\cite{mf2}, a detailed study of 
X-ray surface density $\sigma$- map and the strong and weak gravitational 
lensing convergence $\kappa$-map for the Bullet Cluster has been done and it was 
compared with MOG and dark matter. In ~\cite{mf3}, a critical test of 
MOG  without dark matter and the galaxy rotation velocity curves determined 
observationally which is in excellent agreement with data for the Milky Way 
without a dark matter halo. The observables like shadow cast of non-rotating 
and rotating MOG BH have been studied in ~\cite{mf5}. When the value of MOG 
parameter increses from zero value it was shown that the 
sizes of the shadow cast for these BHs increases significantly. The shadow cast
measured by Event Horizon Telescope~(EHT) confirmed  the result of Einstein's GTR 
whether it is correct or whether it should be modified  under 
strong gravitational fields. 

In~\cite{mf6}, the BHs in MOG has been studied and whether the author derived the 
equations of motion of a test partilcle, stability condition, the radii of circular 
photon orbit and the shadow cast in details. The gravitational lensing 
properties of Kerr-MOG has been studied in ~\cite{mf9} The Kerr-MOG BH merger and 
the ringdown radiation have been considered in ~\cite{liu18}.  The superradiance in 
Kerr-MOG has been examined in~\cite{mf8} very recently.

One aspect that has been never published in the literature is that the computation of epicyclic frequencies 
for the above mentioned BH. It is well known that the circular geodesics of test particles 
are described by three fundamental frequencies: the Keplerian frequency~($\nu_{\phi}$), the radial epicyclic 
frequency~($\nu_{r}$) and the vertical epicyclic frequency~($\nu_{\theta}$). 
In this work, we wish to compute these frequencies for modified gravity which was not 
studied previously. In Newtonian gravity, these characteristic frequencies have the same 
value while in Einstein's gravity they satisfied the inequality: $\nu_{\phi}\geq \nu_{\theta}>\nu_{r}$.

It must be noted that the epicyclic frequencies are key ingredients for the geodesic models of 
quasi-periodic-oscillations~(QPO)~\cite{maselli17}. This QPOs could be help us in a novel way 
to testify the strong gravity. The geodesic models are described by relativistic precession 
model~(RPM)~\cite{stella99} and epicyclic resonance model~(ERM)~\cite{torok05}. Both models 
signal that there exists both low frequency~(LF) QPO and twin high frequency~(HF) QPO. These 
frequencies of QPOs in accreting neutron star  should be measured in near future by 
very-large-area X-ray instrument. The currently available QPO measurement instrument is 
Rossi X-ray Timing Explorer~(RXTE/PCA). The other instruments are eXTP, LOFT or STROBE-X. 
From RPM, it is known that the upper and lower HF QPOs meets with the azimuthal frequency, 
$\nu_{per}=\nu_{\phi}-\nu_{r}$. While the LF QPOs are governed by the nodal precession 
frequency, $\nu_{nod}=\nu_{\phi}-\nu_{\theta}$. These three QPOs signals 
$(\nu_{\phi},\nu_{per}, \nu_{nod})$ yield at the same orbital radius.

The paper has two sections. In first section we have studied the Penrose process for 
Kerr-MOG BH. While in second section, we have computed the epicyclic frequencies for 
circular geodesics. In sub-section 2.1, we have discussed the energy limits on the 
Penrose process followed by the work of Wald. The Bardeen-Press-Teukolsky inequality 
derived in sub-sec. 2.2. In sub-sec. 2.3, we have introduced the concept of
irreducible mass in Kerr-MOG BH. Finally, we have given a brief discussion and outlook in section.~3. 
In Appendix, we have computed the ISCO energy for extremal Kerr-MOG BH.

\section{\label{kmg}  The Penrose Process in Kerr-MOG BH}
Before describing the Penrose process we would like to first describe the basic 
feature of  Kerr-MOG BH. It is an axisymmetric class of spinning BH and it is 
described by the ADM mass parameter~(${\cal M}$), spin parameter~($a$) and a 
deformation parameter or MOG parameter~($\alpha$). This parameter 
$\alpha=\frac{G-G_{N}}{G_{N}}$ should be measured deviation of MOG 
from GR. The basic postulate in MOG theory is that the charge 
parameter is proportional to the square root of the MOG parameter 
i.e. ${\cal Q}=\sqrt{\alpha G_{N}}M$~\cite{mf5}.

The Kerr-MOG BH metric~(in units where $c=1$) can be written in Boyer-Lindquist coordinates 
$(t, r, \theta, \phi)$ as~\cite{mf5}
\begin{eqnarray}
ds^2 =- \frac{\Delta_{r}}{\rho^2} \, \left[dt-a\sin^2\theta d\phi \right]^2+\frac{\sin^2\theta}{\rho^2} \,
\left[(r^2+a^2) \,d\phi-a dt\right]^2
+\rho^2 \, \left[\frac{dr^2}{\Delta_{r}}+d\theta^2\right] ~.\label{mg2.1}
\end{eqnarray}
where
\begin{eqnarray}
\rho^2 & \equiv & r^2+a^2\cos^2~\theta \nonumber\\
\Delta_{r} & \equiv & r^2-2G_{N}(1+\alpha)Mr+a^2 + G_{N}^2 \alpha(1+\alpha) M^2  ~.\label{m2.1}
\end{eqnarray}
where $G_{N}$ is a Newtonian constant and $M$ is the Komar mass~\cite{mf8}. 
For simplicity, we have taken the value of $G_{N}=1$ hereafter and througout the 
work. The ADM mass and angular momentum computed in~\cite{ps} as ${\cal M}=(1+\alpha)M$ and $J=a{\cal M}$
~\footnote{We find the relation between the Komar mass and ADM mass is $M=\frac{\cal M}{1+\alpha}$. If one can 
consider either the Komar mass or the ADM mass in the calculation then the physics will not be change. We here 
consider the ADM mass througout the calculation for convenience. }. 
Substituting these values in Eq.~(\ref{m2.1}) $\Delta_{r}$ becomes
\begin{eqnarray}
\Delta_{r} & = & r^2-2{\cal M}r+a^2 + \frac{\alpha}{(1+\alpha)} {\cal M}^2 
\end{eqnarray}

\begin{figure}
\begin{center}
{\includegraphics[width=0.45\textwidth]{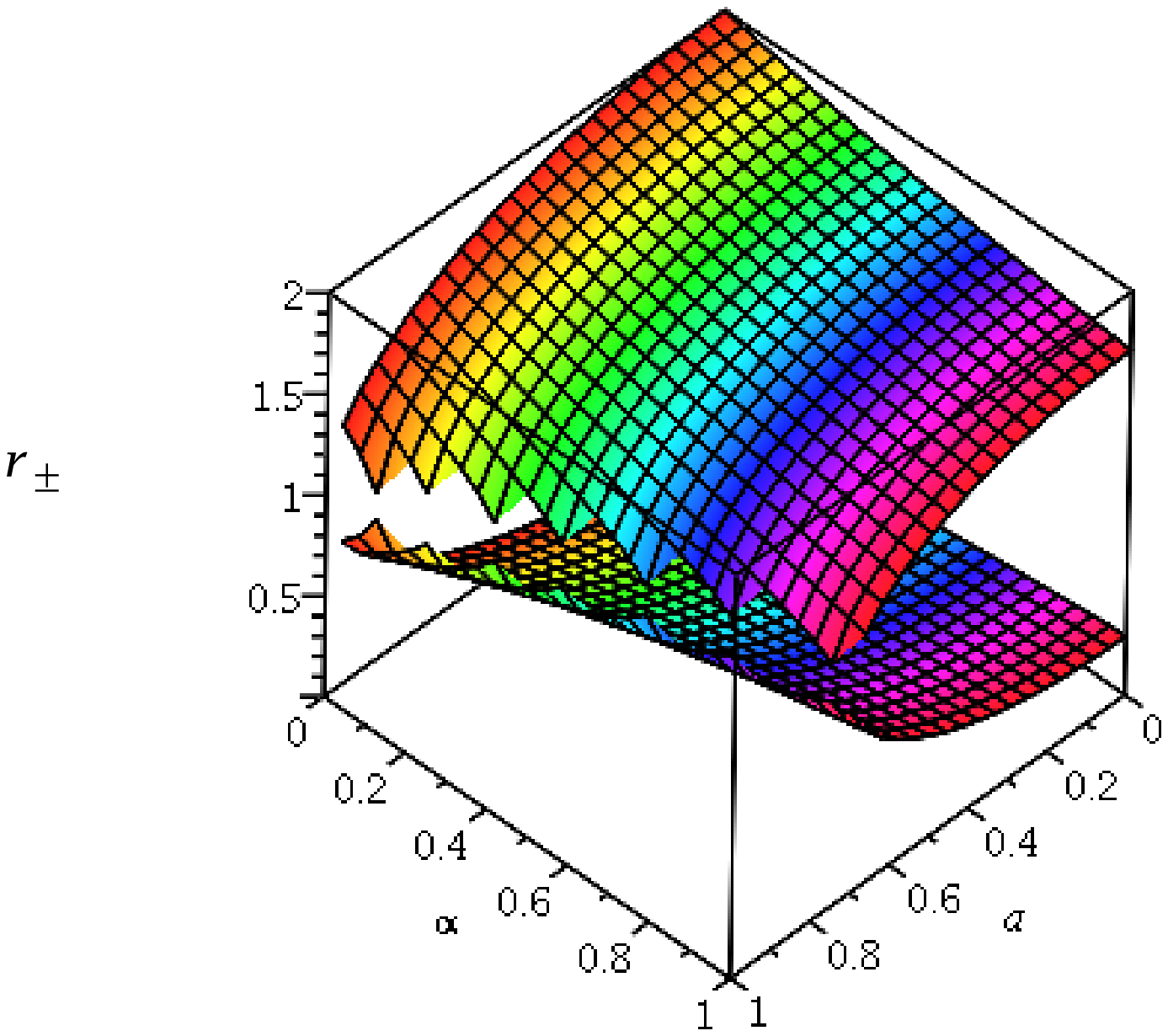}}
{\includegraphics[width=0.45\textwidth]{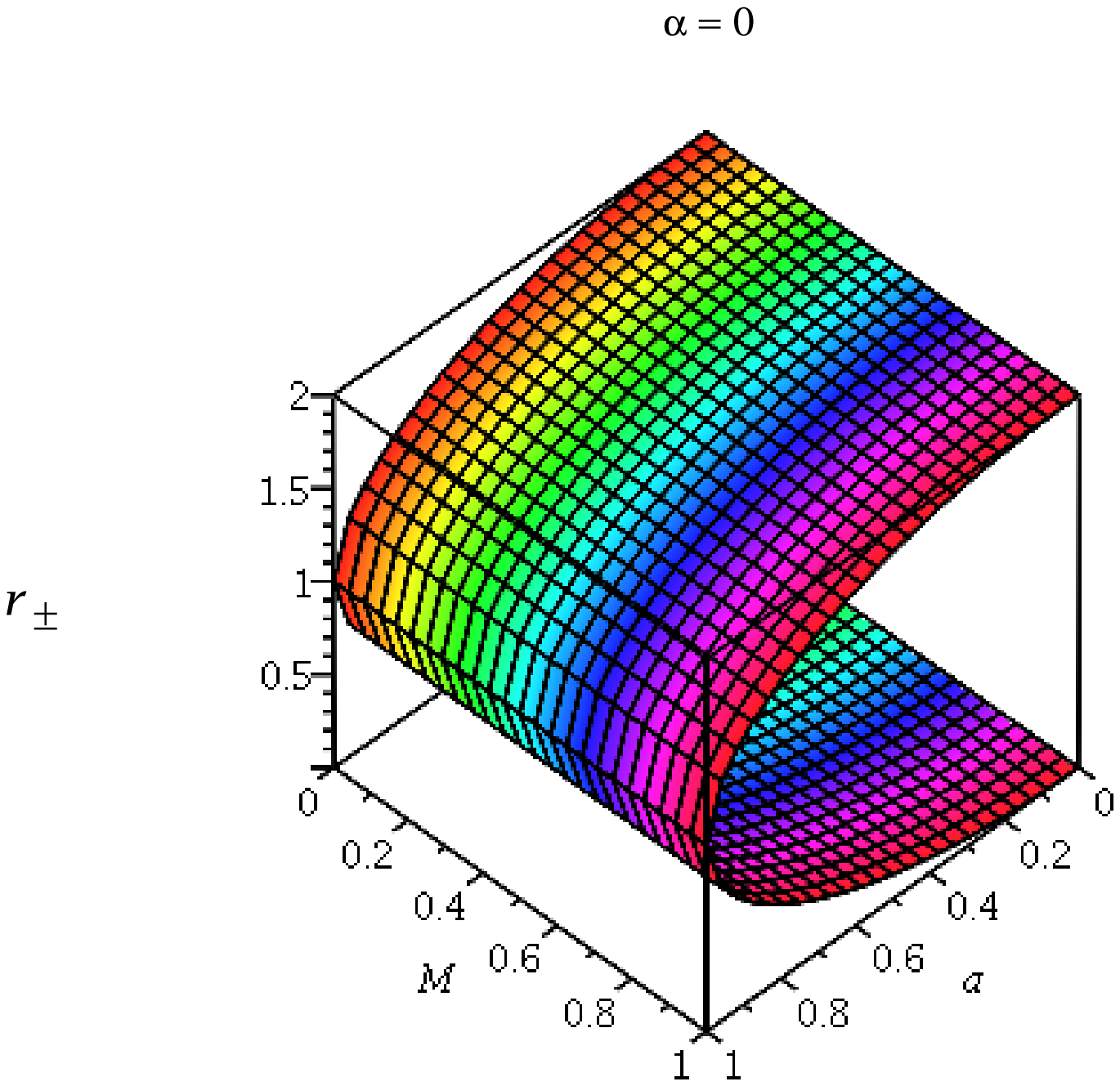}}
{\includegraphics[width=0.45\textwidth]{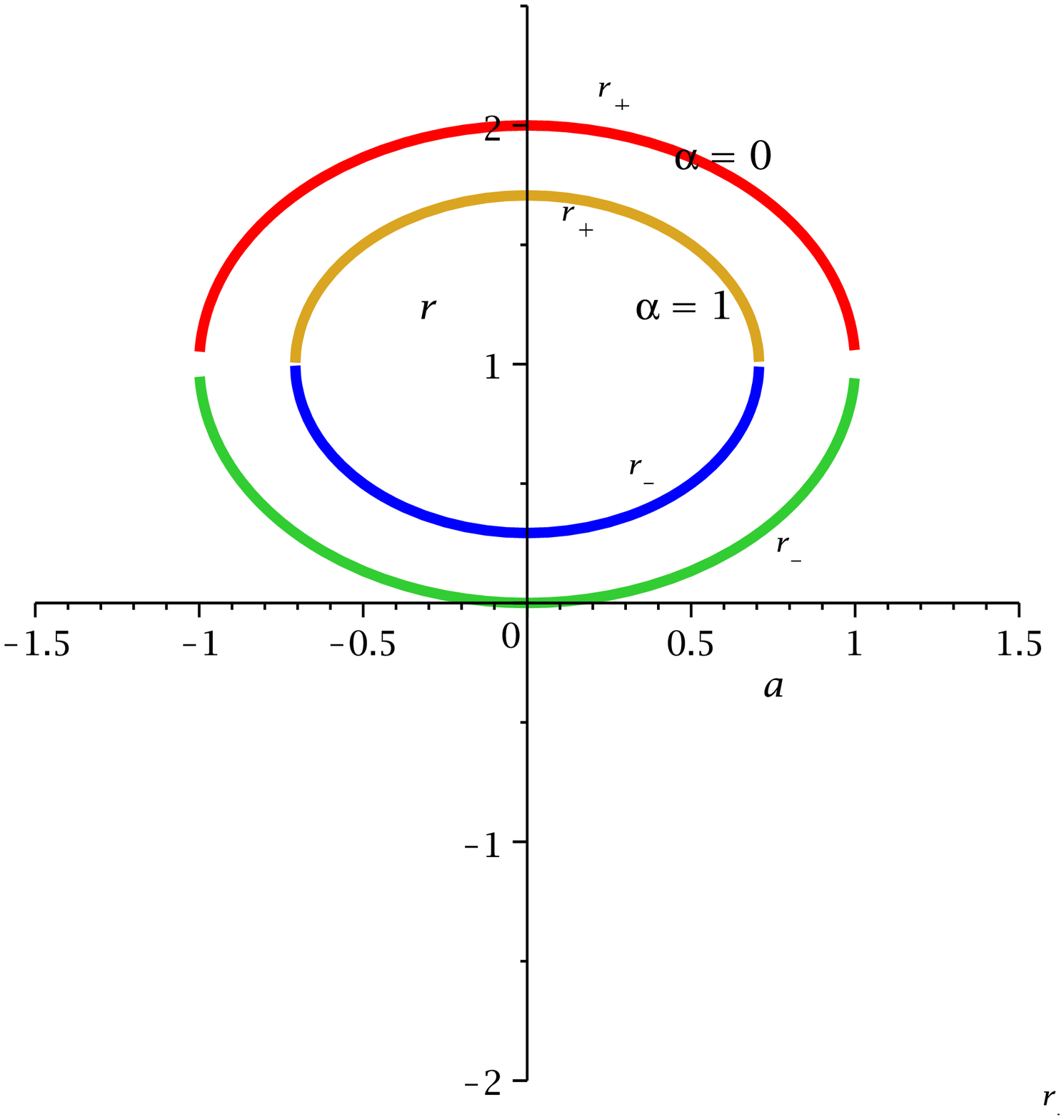}}
\end{center}
\caption{The figure shows the variation  of $r_{\pm}$  with $a$ and $\alpha$ for Kerr BH
and Kerr-MOG BH.
\label{mvf}}
\end{figure}

The BH consists of two horizons namely  event horizon~($r_{+}$) and Cauchy horizon~($r_{+}$). They are denoted as 
\begin{eqnarray}
r_{\pm} &=&  {\cal M} \pm \sqrt{\frac{{\cal M}^2}{1+\alpha}-a^2} ~. \label{mg2.2}
\end{eqnarray}
It may be noted that when $\alpha=0$, one obtains the horizon radii of Kerr BH. The BH solution exists 
when $\frac{{\cal M}^2}{1+\alpha} > a^2$. When $\frac{{\cal M}^2}{1+\alpha}=a^2$, one finds the 
extremal BH. When $\frac{{\cal M}^2}{1+\alpha}<a^2$, one obtains the naked singularity case. The behavior 
of the outer horizon and inner horizon could be found in the Fig.~\ref{mvf}. It follows from the figure that 
the presence of the MOG parameter could somehow deformed the shape of the horizon radii. 
The ergosphere is situated  at 
\begin{eqnarray}
r &=& r_{e} (\theta) = {\cal M} + \sqrt{\frac{{\cal M}^2}{1+\alpha}-a^2 \cos^2~\theta} ~. \label{mg2.3}
\end{eqnarray}
This surface is outer to the event horizon  and it coincides with event horizon at 
the poles $\theta=0$ and $\theta=\pi$. 
\begin{figure}
\begin{center}
{\includegraphics[width=0.45\textwidth]{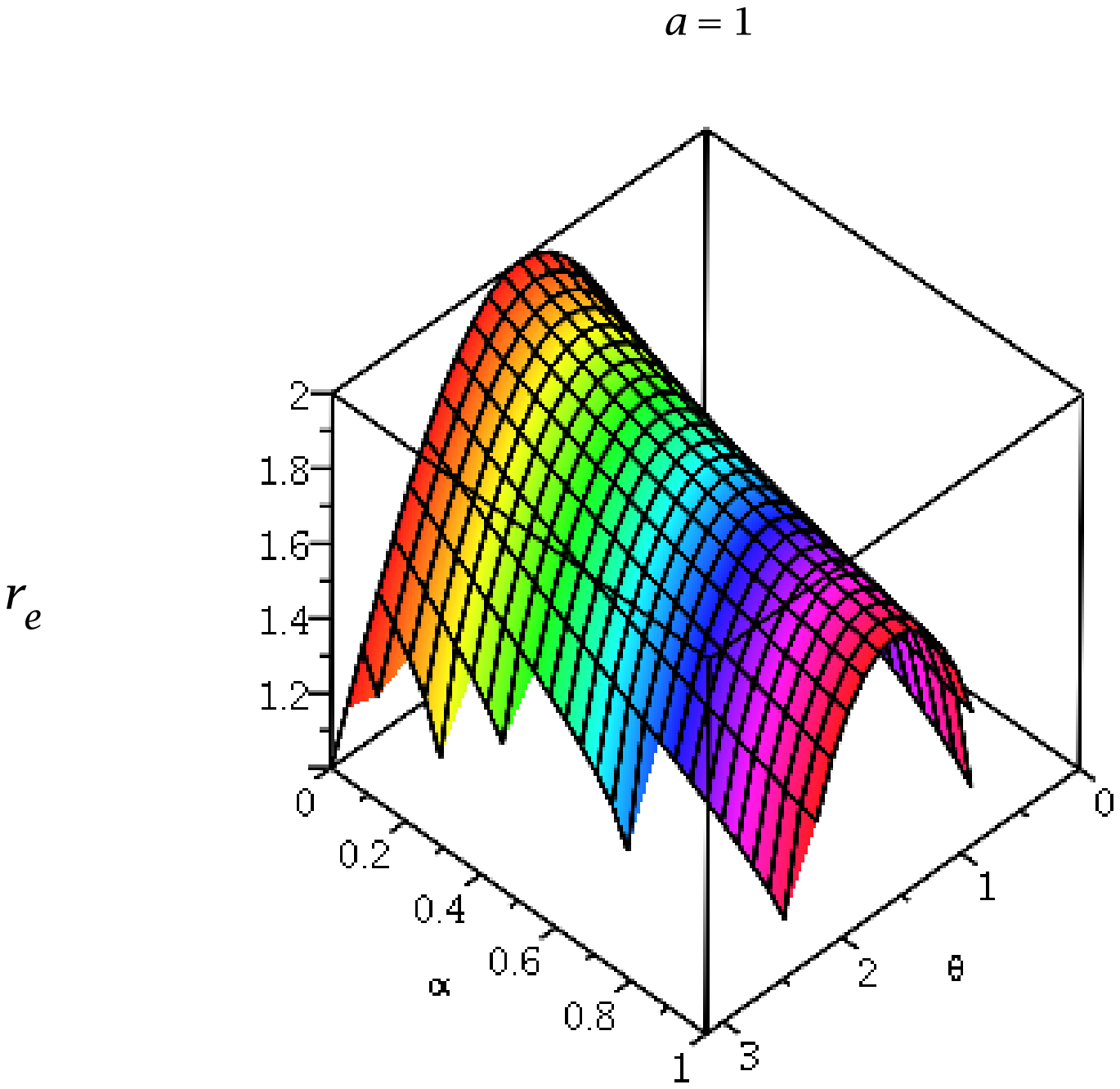}}
{\includegraphics[width=0.45\textwidth]{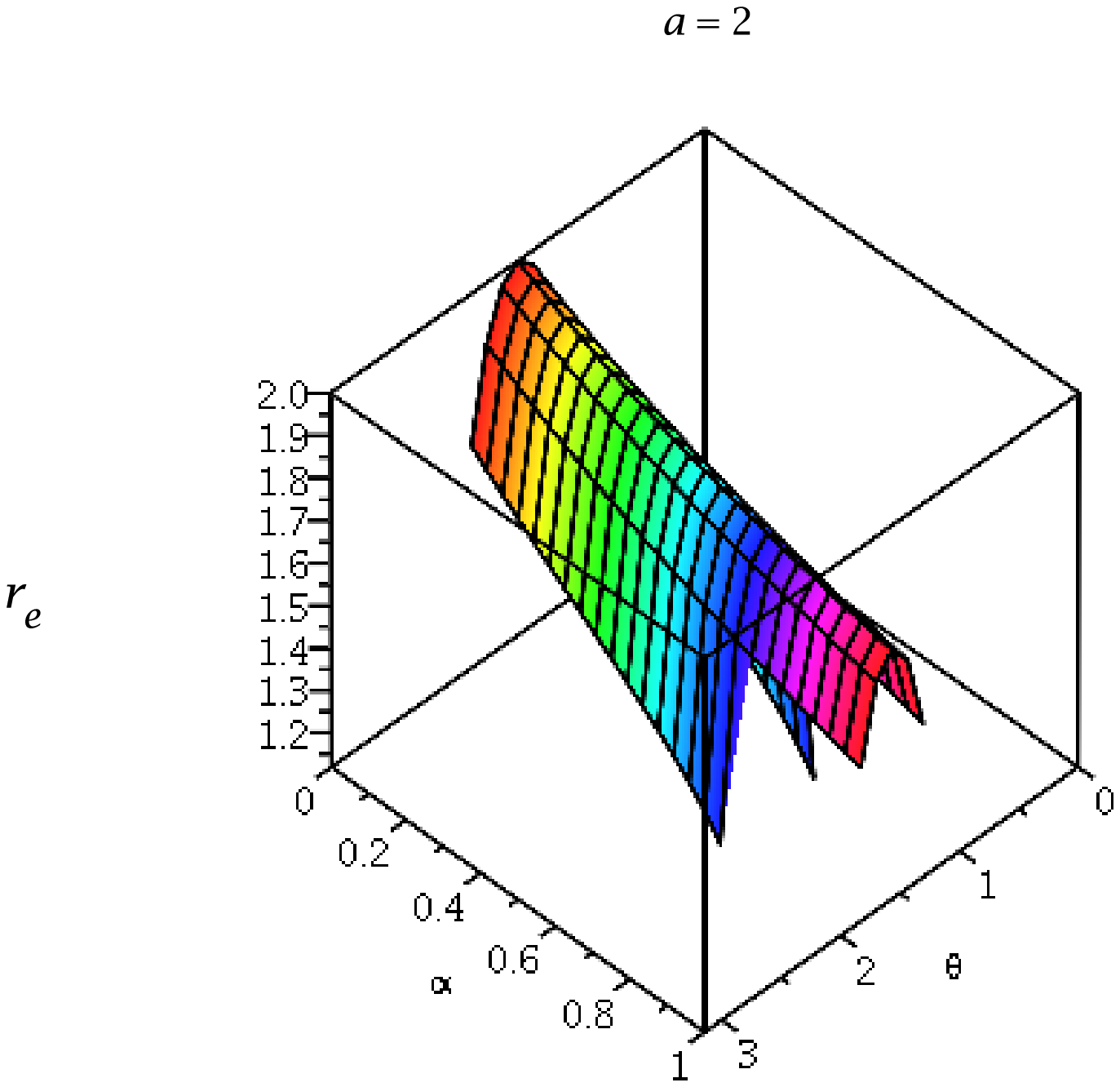}}
{\includegraphics[width=0.45\textwidth]{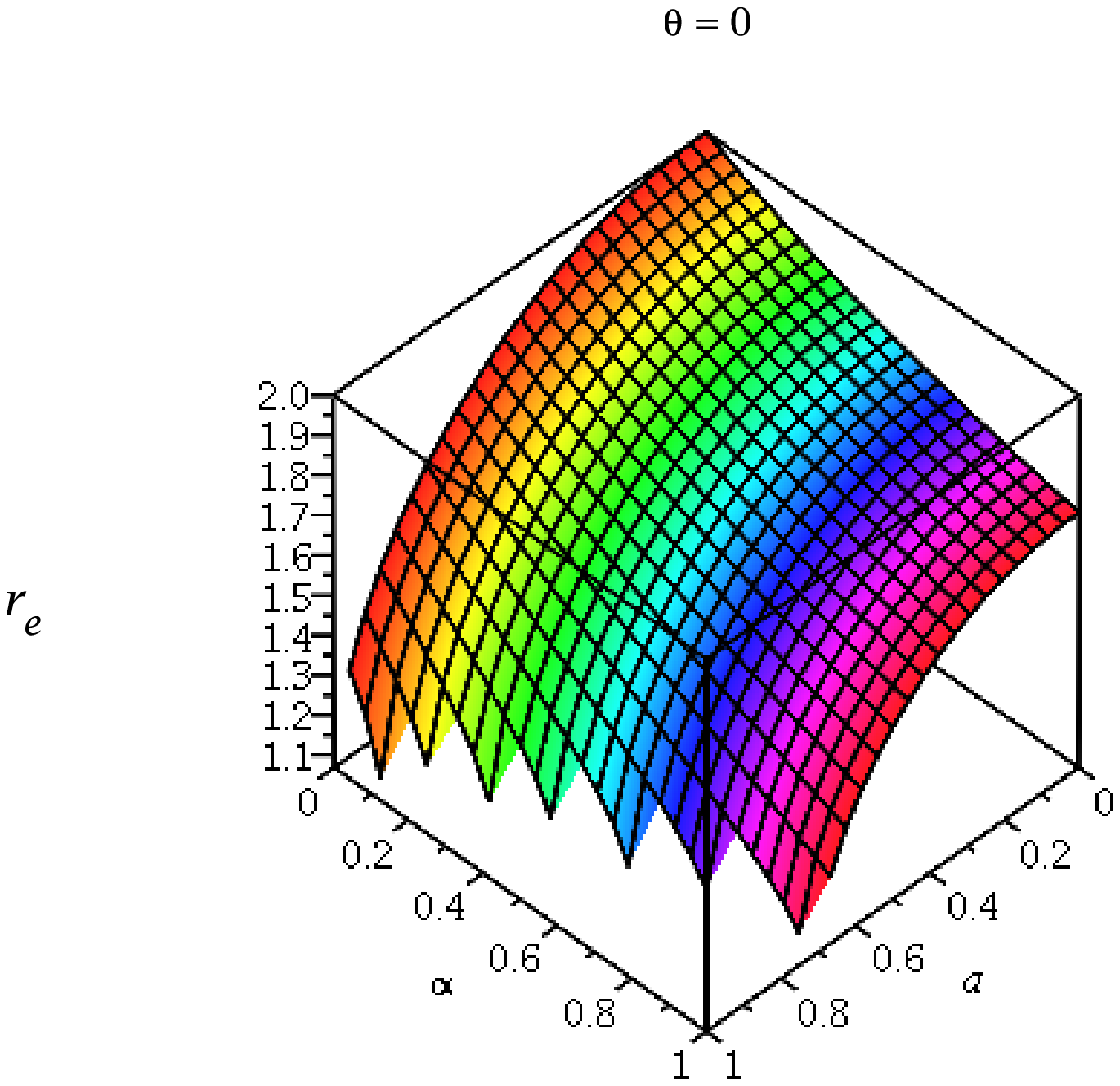}}
{\includegraphics[width=0.45\textwidth]{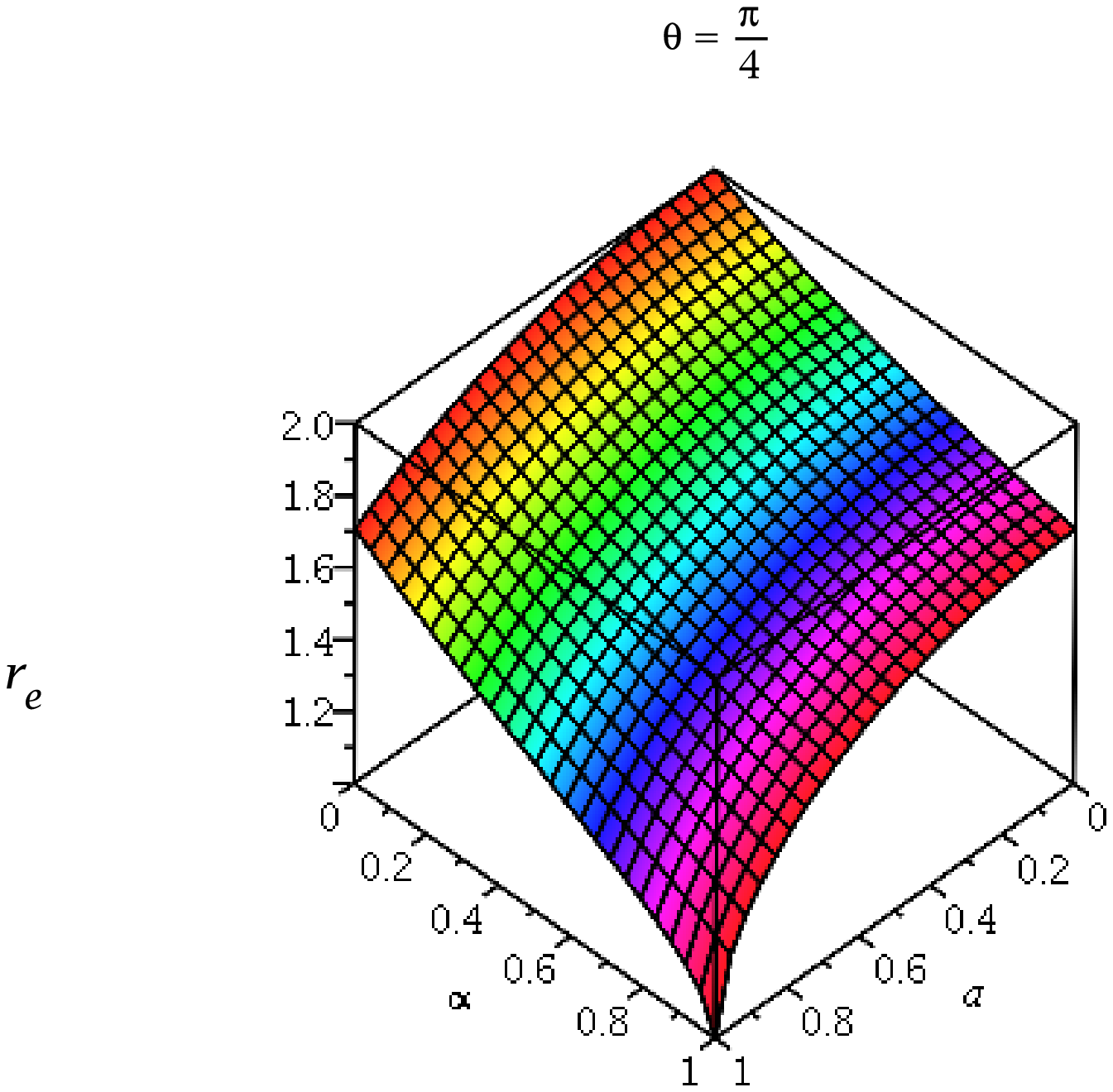}}
{\includegraphics[width=0.45\textwidth]{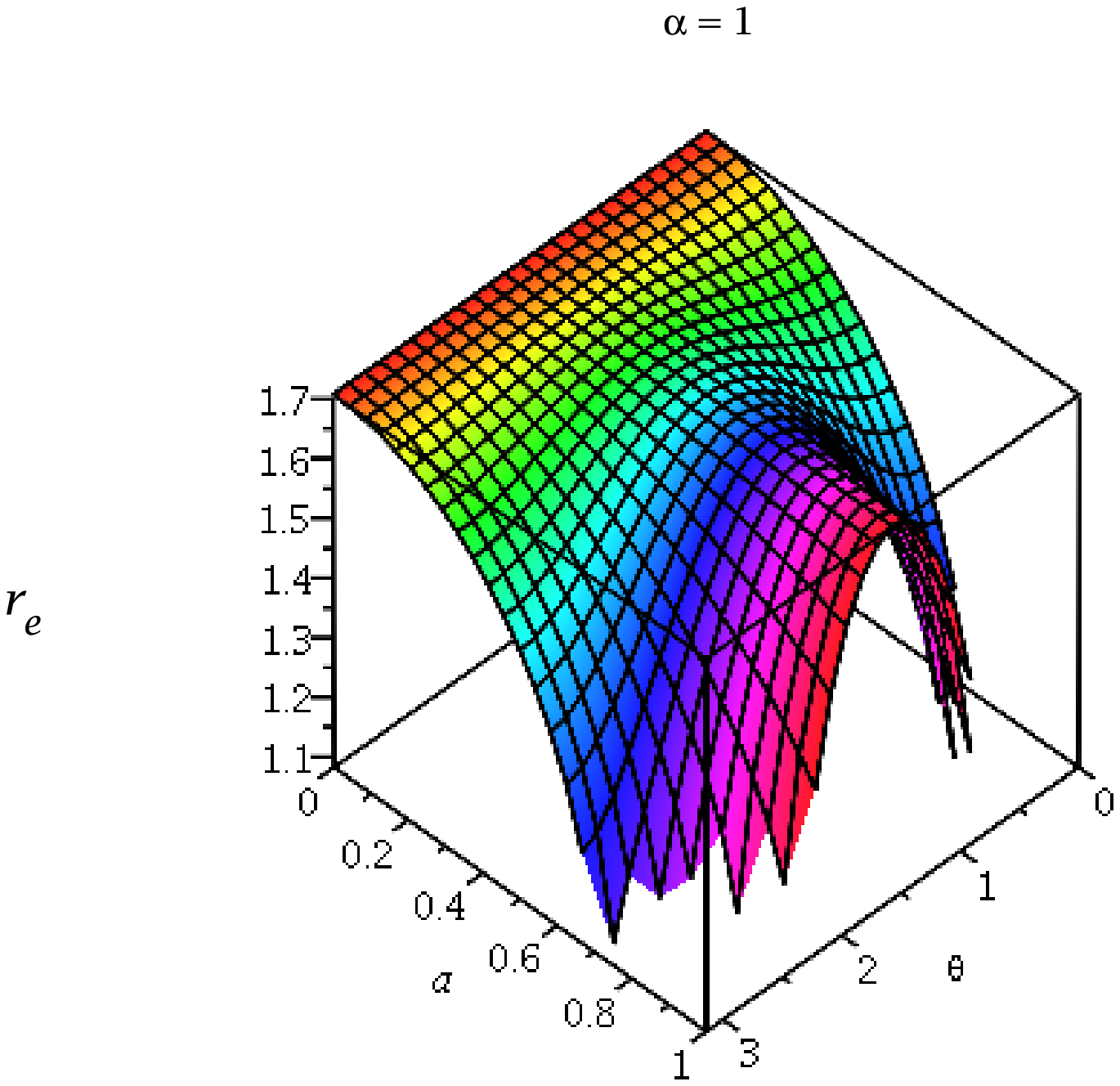}}
{\includegraphics[width=0.45\textwidth]{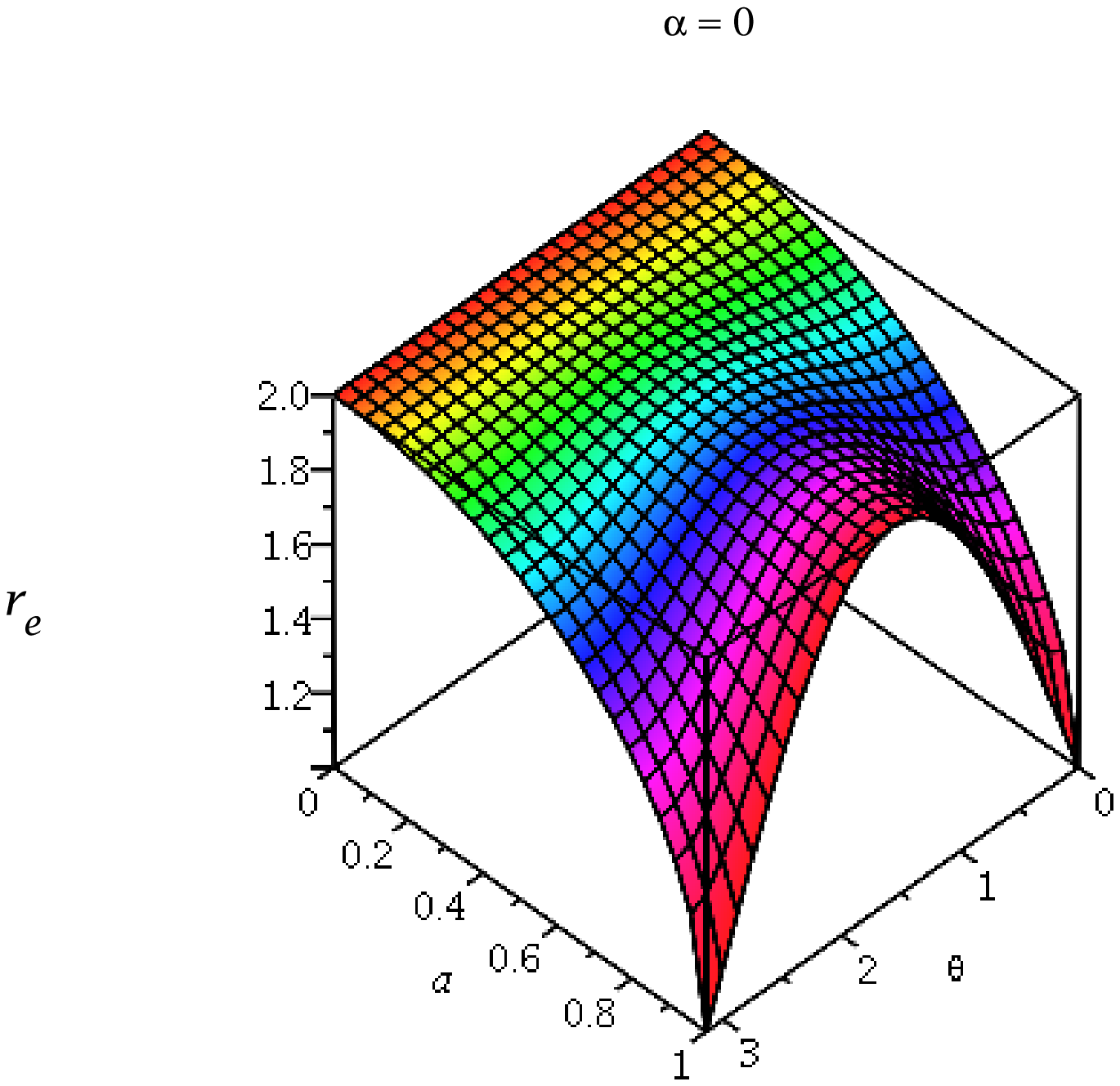}}
\end{center}
\caption{The figure shows the variation  of $r_{e}$  with $a$ and $\alpha$ for Kerr BH
and Kerr-MOG BH.
\label{mve}}
\end{figure}
To obtain the radial equation for the geodesic motion of a test particle in Kerr-MOG BH, we have followed 
the book of S. Chandrashekar \cite{sch}. We should also restricted in the equatorial plane. 
Therefore the Lagrangian density for the geodesic motion of a test particle could be written as
$$
2{\cal L} =- \left(1-\frac{2{\cal M}}{r}+\frac{\alpha}{1+\alpha} \frac{{\cal M}^2}{r^2}\right)\,{\dot{t}}^2
-2a\left(\frac{2{\cal M}}{r} -\frac{\alpha}{1+\alpha} \frac{{\cal M}^2}{r^2}\right)\,\dot{t}\,\dot{\phi}+
$$
\begin{eqnarray}
\frac{r^2}{\Delta_{r}}\,{\dot{r}}^2+\left(r^2+a^2+\frac{2{\cal M}a^2}{r}-
\frac{\alpha}{1+\alpha} \frac{a^2{\cal M}^2}{r^2}\right) \,{\dot{\phi}}^2 ~.\label{mlag}
\end{eqnarray}
The radial equation that governs the geodesic structure of Kerr-MOG BH is given by  
$$
\dot{r}^{2} = {\cal E}^2\left(1+\frac{a^2}{r^2}+\frac{2{\cal M}a^2}{r^3}-\frac{\alpha}{1+\alpha} 
\frac{a^2{\cal M}^2}{r^4}\right)
$$
\begin{eqnarray}
-\frac{\ell^2}{r^2}\left(1-\frac{2{\cal M}}{r}+\frac{\alpha}{1+\alpha} \frac{{\cal M}^2}{r^2}\right)
-2a \ell {\cal E} \left(\frac{2{\cal M}}{r^3} -\frac{\alpha}{1+\alpha} \frac{{\cal M}^2}{r^4} \right) 
+\epsilon \frac{\Delta_{r}}{r^2} ~.\label{mradial}
\end{eqnarray}
where  $\epsilon=-1$ for time-like geodesics and $\epsilon=0$ for null geodesics.  Also, ${\cal E}$ corresponds 
to the energy and $\ell$ corresponds to the angular momentum of the test particle.

To study the Penrose process one should use the radial geodesic equation i.e. 
Eq.~(\ref{mradial}) then
$$
{\cal E}^2\left(r^4+a^2r^2+2{\cal M}a^2r-\frac{\alpha}{1+\alpha}{\cal M}^2a^2\right)-
2 a{\cal E}\ell\left(2{\cal M} r-\frac{\alpha}{1+\alpha}{\cal M}^2\right)
$$
\begin{eqnarray}
-\ell^2\left(r^2-2{\cal M}r+\frac{\alpha}{1+\alpha}{\cal M}^2\right)+\epsilon \Delta_{r} r^2 &=& 0 ~.\label{mradpp}
\end{eqnarray}
Since there is no contribution to ${\cal E}$ from the kinetic energy part hence one could solve the above 
equation for both  ${\cal E}$ and $\ell$ as separately then

\begin{eqnarray}
{\cal E} =\frac{a \ell \left(2{\cal M}r-\frac{\alpha}{1+\alpha}{\cal M}^2 \right) \pm Z_{r} \sqrt{\Delta_{r}}}
{r^4+a^2r^2+2{\cal M}a^2r-\frac{\alpha}{1+\alpha}{\cal M}^2a^2} ~.\label{mengpp}
\end{eqnarray}
where
$$
Z_{r} =\sqrt{\ell^2 r^4-\epsilon r^2 \left[r^4+a^2\left(r^2+2{\cal M}r-\frac{\alpha}{1+\alpha}{\cal M}^2 \right)\right]}
$$
and
\begin{eqnarray}
\ell &=& \frac{-a{\cal E} \left(2{\cal M}r-\frac{\alpha}{1+\alpha}{\cal M}^2 \right) \pm U_{r} 
\sqrt{\Delta_{r}}}{r^2-2{\cal M}r+\frac{\alpha}{1+\alpha}{\cal M}^2} ~.\label{mangpp}
\end{eqnarray}
where
$$
U_{r} = \sqrt{{\cal E}^2r^4+ \epsilon r^2\left(r^2-2{\cal M}r+\frac{\alpha}{1+\alpha}{\cal M}^2\right)}
$$

The above equations have been derived using the following important identity
$$
r^4 \Delta_{r}-a^2 \left(2{\cal M}r-\frac{\alpha}{1+\alpha}{\cal M}^2 \right)^{2} =
$$
\begin{eqnarray}
\left(r^4+a^2r^2+2{\cal M}a^2r-\frac{\alpha}{1+\alpha}{\cal M}^2 a^2 \right)
\left(r^2-2{\cal M}r+\frac{\alpha}{1+\alpha}{\cal M}^2\right)~.\label{midentitypp}
\end{eqnarray}
Using Eq.~(\ref{mengpp}), one could derive the condition while the value of the energy is 
negative as discerned by an observer at infinity. With out loss of generality we 
have taken the value of ${\cal E}=1$ when a particle of unit mass, at rest at infinity. 
Therefore at the present moment we have considered the positive sign in the right hand 
side of the Eq.~(\ref{mengpp}). Thus it must be obeyed that the following criterion should be 
satisfied for 
${\cal E}<0$, $\ell<0$ and
$$
a^{2}\ell^{2} \left(2{\cal M}r-\frac{\alpha}{1+\alpha}{\cal M}^2 \right)^2 >
$$
\begin{eqnarray}
\Delta_{r} \, r^2 \left[\ell^2 r^4-\epsilon r^2\left(r^4+a^2r^2+2{\cal M}a^2r-
\frac{\alpha}{1+\alpha}{\cal M}^2 a^2 \right)\right] 
~.\label{mien}
\end{eqnarray}
Using Eq.~(\ref{midentitypp}), the above inequality could be written as 
$$
\left(r^4+a^2r^2+2{\cal M}a^2r-\frac{\alpha}{1+\alpha}{\cal M}^2 a^2 \right) \times
$$
\begin{eqnarray}
\left[\ell^2 \left(r^2-2{\cal M}r+\frac{\alpha}{1+\alpha}{\cal M}^2\right)-\epsilon \Delta_{r} r^2 \right] < 0
~.\label{midenpp}
\end{eqnarray}
It immediately suggests that ${\cal E}<0$ if and only if $\ell<0$. Also and
\begin{eqnarray}
\left(1-\frac{2{\cal M}}{r}+\frac{\alpha}{1+\alpha}\frac{{\cal M}^2}{r^2}\right)<\frac{\Delta_{r}}{\ell^2} \epsilon
\end{eqnarray}
Therefore the only possibility in the equatorial plane is that the counter-rotating particles should 
have negative energy and it happens inside the ergosphere. This ergosphere radius for Kerr-MOG BH has 
been given in Eq.~(\ref{mg2.3}). For extremal Kerr-MOG BH, the ergo-sphere occurs at 
$r_{e} (\theta) = {\cal M} + a \sin\theta$ which is exactly same as the ergosphere 
radius of extreme Kerr BH.

What exactly happens in this process is that when a particle at rest at infinity arrives 
at a point $r<a+M$ in the equatorial plane it has a turning point in such a way that 
$\dot{r}=0$. At the meeting point $r$, the particle splits into two photons: one photon 
crosses the event horizon and is lost when the other one escapes to infinity. We could arrange 
this process in such a way that the photon which crosses the event horizon has negative energy 
and the photon which escapes to infinity has more energy than the particle which arrived from 
infinity.

Now let us suppose ${\cal E}^{(x)}=1,\, \ell^{(x)}; \, {\cal E}^{(y)},\, \ell^{(y)};
\,\, \mbox{and}\,\, {\cal E}^{(z)}, \ell^{(z)}$
are the energies and the angular momentum of the particle arriving from
infinity and of the photons which cross the outer horizon  and escape
to infinity, respectively.

Since the particles come from infinity and get at $r$ followed by a time-like 
circular geodesics then it has a turning point at $r$, its angular momentum, $\ell^{(x)}$, 
could be determined from Eq.~(\ref{mangpp}) by putting $\epsilon=-1, {\cal E}=1$. 
Therefore one gets,
\begin{eqnarray}
\ell^{(x)} &=& \frac{\left[-a\left(2{\cal M}r-\frac{\alpha}{1+\alpha}{\cal M}^2 \right)+r\sqrt{\Delta_{r}}
\sqrt{2{\cal M}r-\frac{\alpha}{1+\alpha}{\cal M}^2 }\right]}
{\left(r^2-2{\cal M}r+\frac{\alpha}{1+\alpha}{\cal M}^2\right)} \nonumber\\
        &=& \chi^{(x)} \, \mbox{(say)} ~.\label{mangppx}
\end{eqnarray}
Similarly, substituting the value of $\epsilon=0$ in Eq.~(\ref{mangpp}) one would get the 
relation between the energy and the angular momenta of the photon which crosses the event 
horizon and the photon which escapes to infinity as
\begin{eqnarray}
\ell^{(y)} &=& \frac{\left[-a\left(2{\cal M}r-\frac{\alpha}{1+\alpha}{\cal M}^2\right) {\cal E}^{(y)}-\sqrt{\Delta_{r}} 
r^2{\cal E}^{(y)} \right]}{\left(r^2-2{\cal M}r+\frac{\alpha}{1+\alpha}{\cal M}^2\right)}\nonumber\\
&=& \chi^{(y)} {\cal E}^{(y)}\, \mbox{(say)} ~.\label{mangppy}
\end{eqnarray}
and
\begin{eqnarray}
\ell^{(z)} &=& \frac{\left[-a\left(2{\cal M}r-\frac{\alpha}{1+\alpha}{\cal M}^2\right){\cal E}^{(z)}
+\sqrt{\Delta_{r}}r^2{\cal E}^{(z)} \right]}{\left(r^2-2{\cal M}r+\frac{\alpha}{1+\alpha}{\cal M}^2\right)}
\nonumber\\
&=& \chi^{z} {\cal E}^{(z)}\, \mbox{(say)} ~.\label{mangppz}
\end{eqnarray}
Now the conservation of energy and angular momentum gives us
\begin{eqnarray}
{\cal E}^{(y)}+{\cal E}^{(z)} &=& {\cal E}^{(x)} = 1
\end{eqnarray}
and
\begin{eqnarray}
\ell^{(y)}+\ell^{(z)} = \chi^{(y)} {\cal E}^{(y)}+\chi^{(z)} {\cal E}^{(z)}= \ell^{(x)}
=\chi^{(x)}
\end{eqnarray}
After solving the above equations, we find
\begin{eqnarray}
{\cal E}^{(y)} &=& \frac{\chi^{(x)} -\chi^{(z)}}{\chi^{(y)}-\chi^{(z)}}
\end{eqnarray}
and
\begin{eqnarray}
{\cal E}^{(z)} &=& \frac{\chi^{(y)} -\chi^{(x)}}{\chi^{(y)}-\chi^{(z)}}
\end{eqnarray}
Putting the values of $\chi^{(x)}$, $\chi^{(y)}$, and $\chi^{(z)}$
by using Eqns.~(\ref{mangppx}), (\ref{mangppz}), we find
\begin{eqnarray}
{\cal E}^{(y)} &=& -\frac{1}{2}\left(\frac{\sqrt{2{\cal M}r-\frac{\alpha}{1+\alpha}{\cal M}^2}}{r}-1 \right)
\end{eqnarray}
and
\begin{eqnarray}
{\cal E}^{(z)} &=& +\frac{1}{2}\left(\frac{\sqrt{2{\cal M}r-\frac{\alpha}{1+\alpha}{\cal M}^2}}{r}-1 \right)
\end{eqnarray}
In the limit $\alpha=0$, one obtains the energy value for Kerr BH. 

\begin{figure}
\begin{center}
{\includegraphics[width=0.45\textwidth]{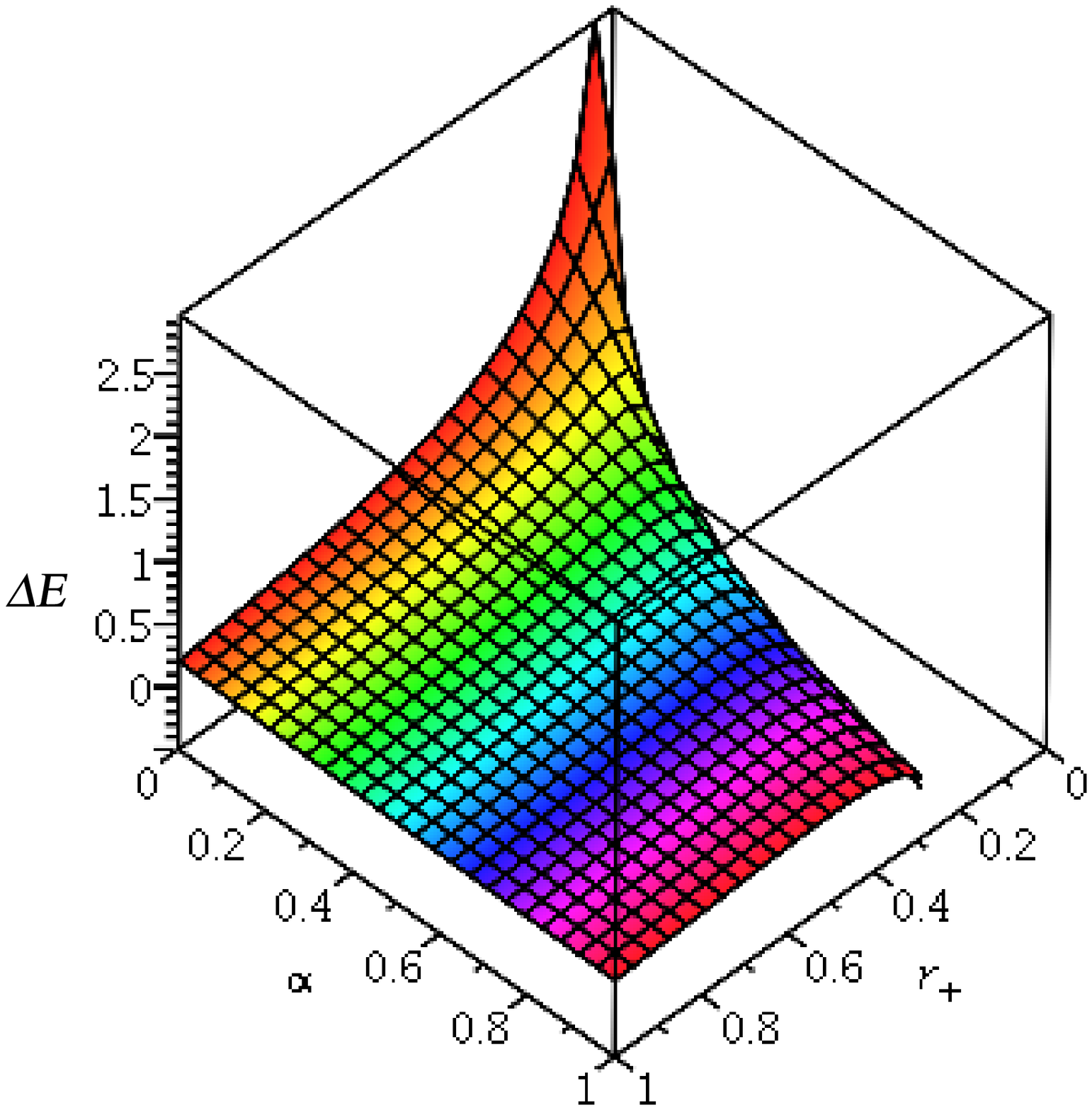}}
{\includegraphics[width=0.45\textwidth]{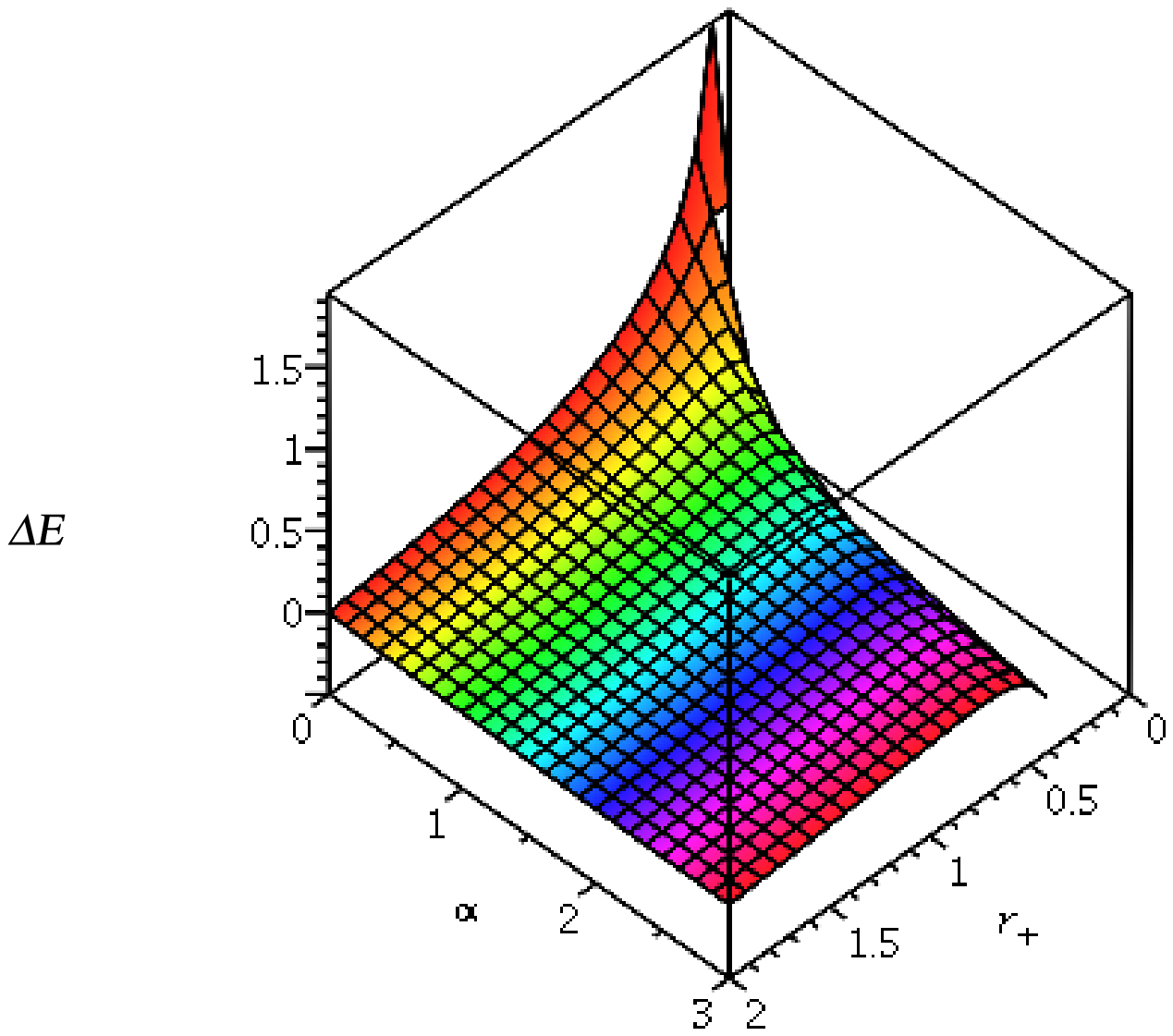}}
{\includegraphics[width=0.45\textwidth]{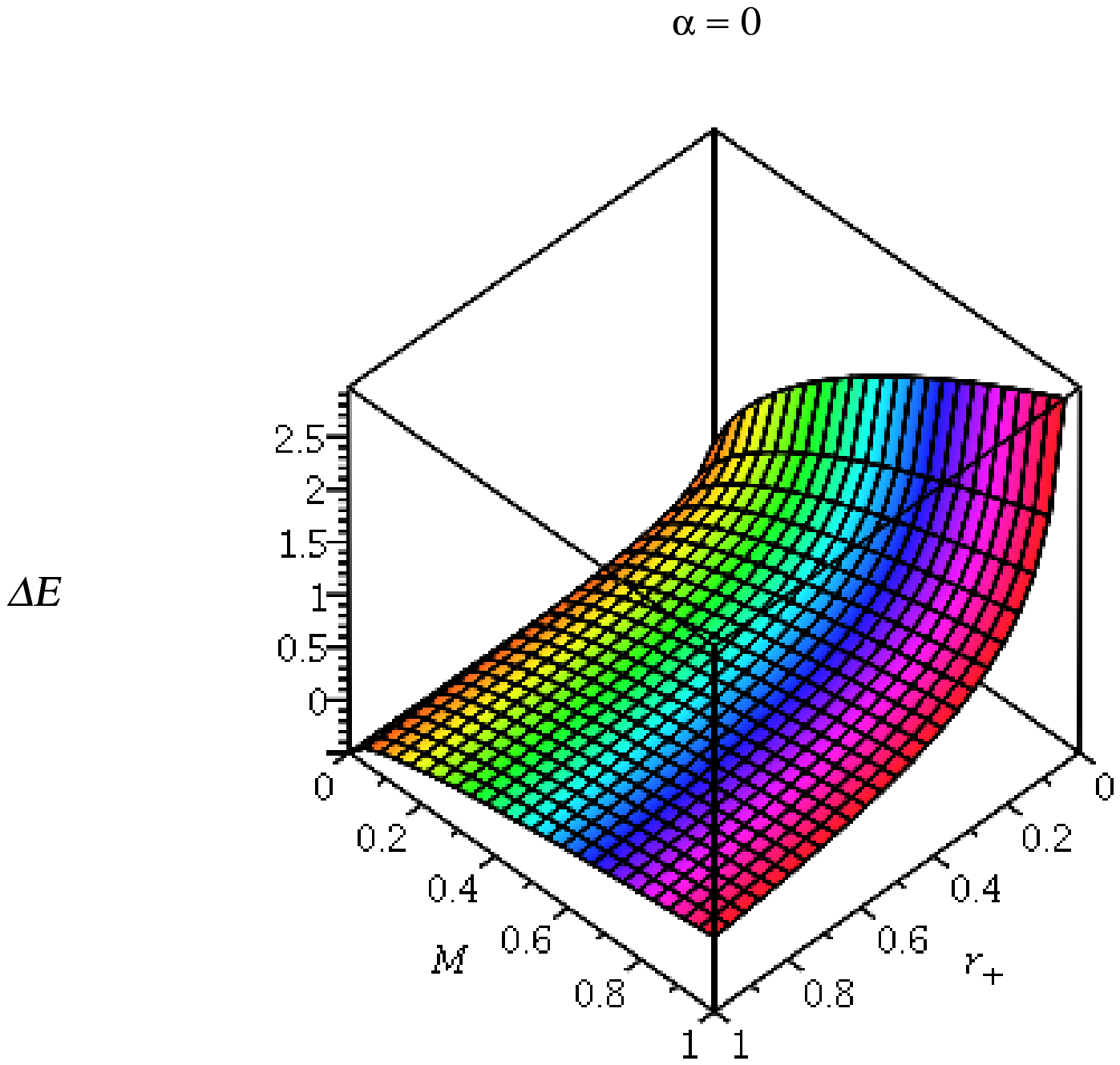}}
{\includegraphics[width=0.45\textwidth]{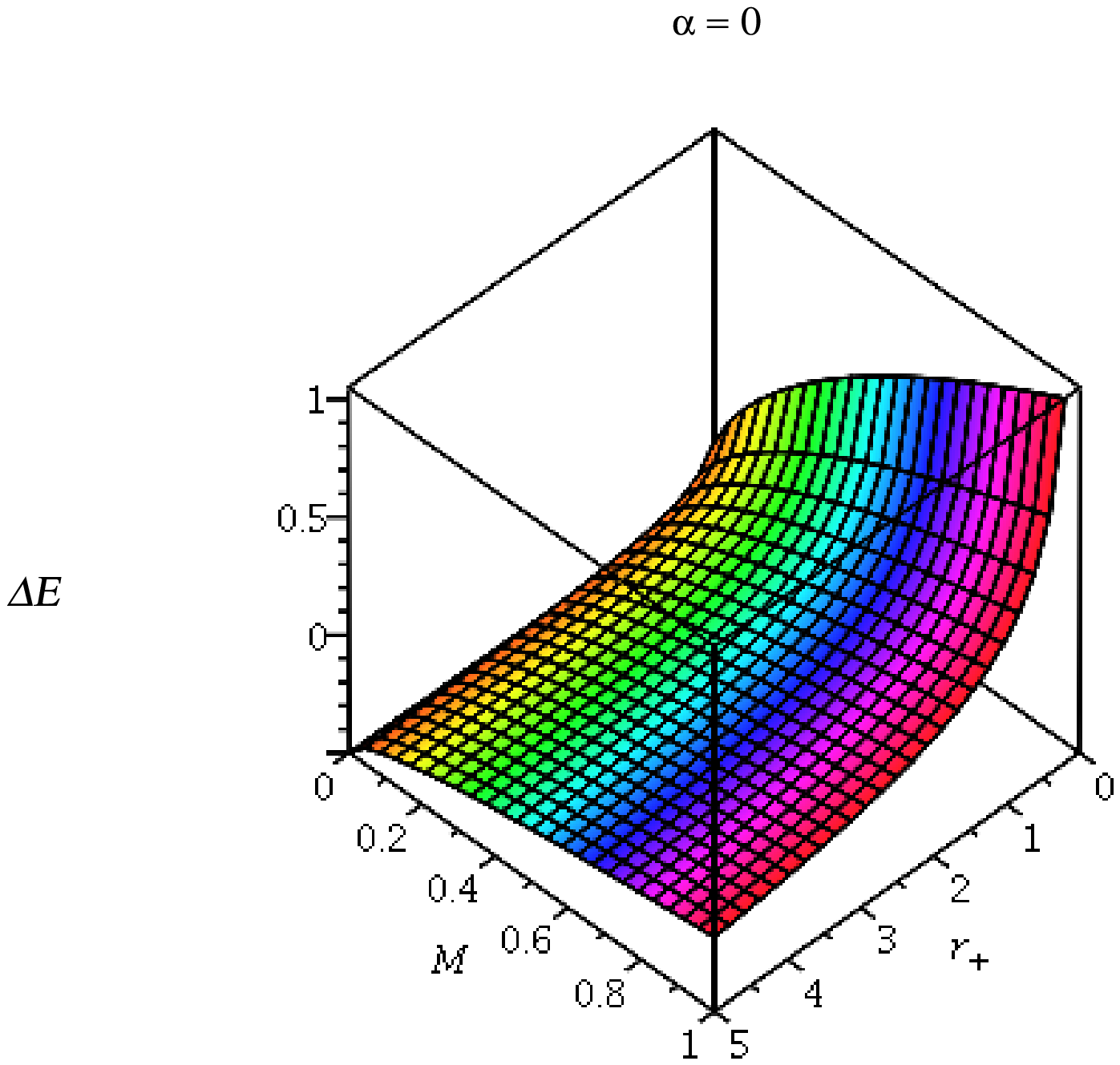}}
\end{center}
\caption{The figure shows the variation  of $\Delta {\cal E}$  with $r_{+}$ for Kerr BH and Kerr-MOG BH.
\label{vmf}}
\end{figure}

\begin{figure}
\begin{center}
{\includegraphics[width=0.45\textwidth]{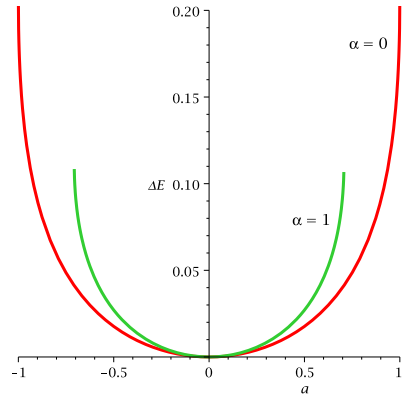}}
\end{center}
\caption{The figure shows the variation  of $\Delta {\cal E}$  with $a$ and $\alpha$, and without $\alpha$.
\label{vfe1}}
\end{figure}

\begin{figure}
\begin{center}
{\includegraphics[width=0.45\textwidth]{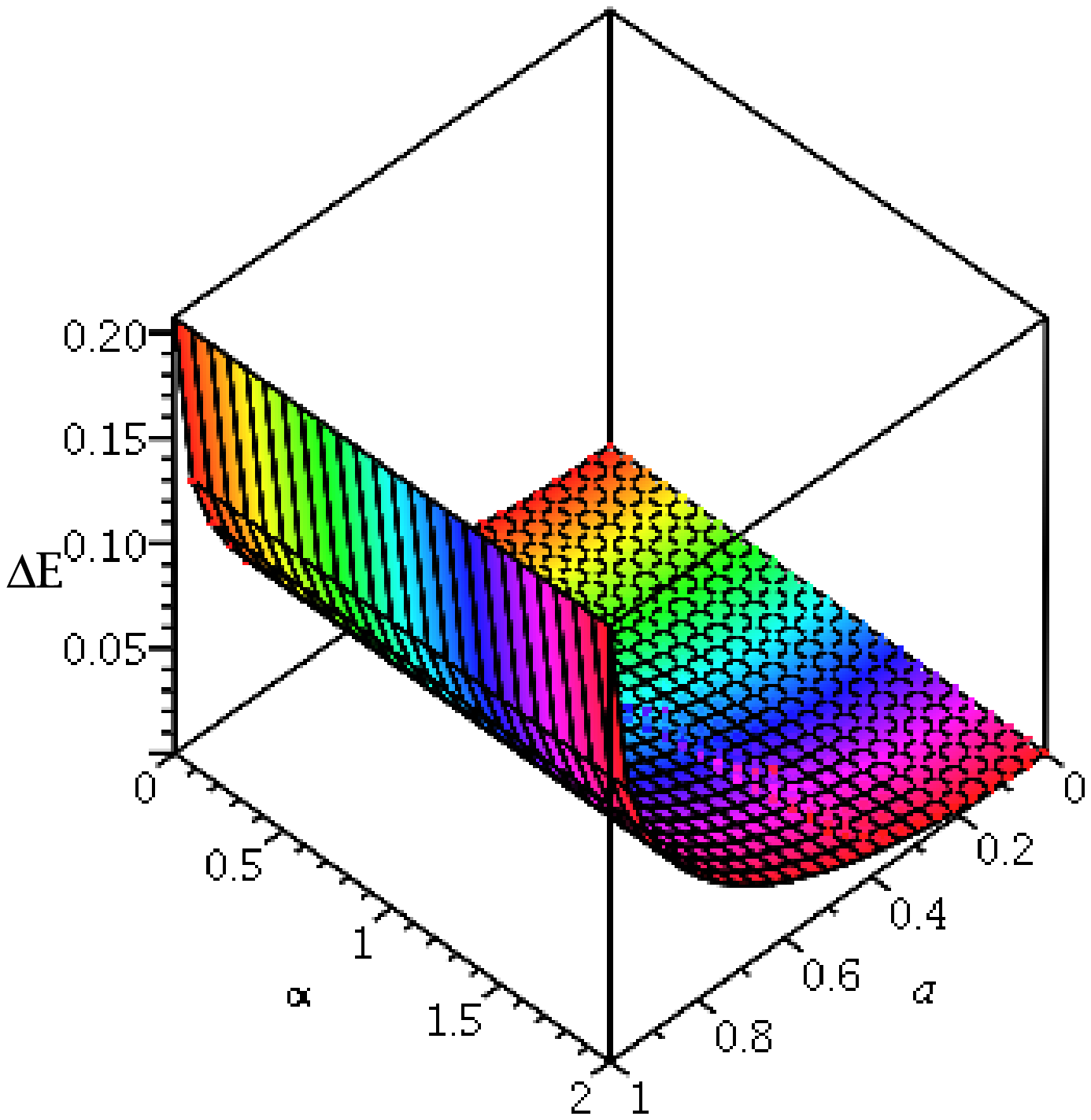}}
{\includegraphics[width=0.45\textwidth]{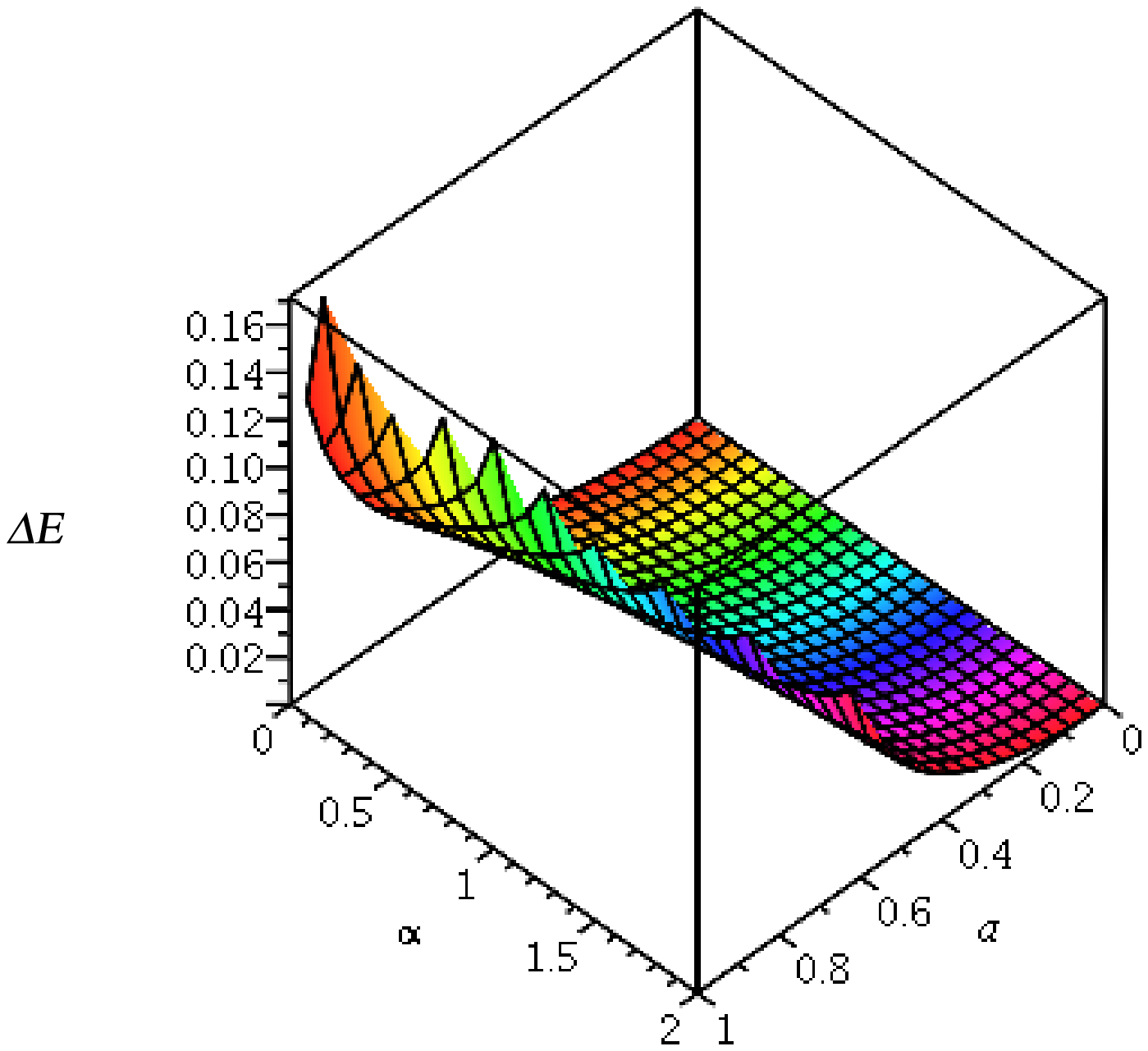}}
{\includegraphics[width=0.45\textwidth]{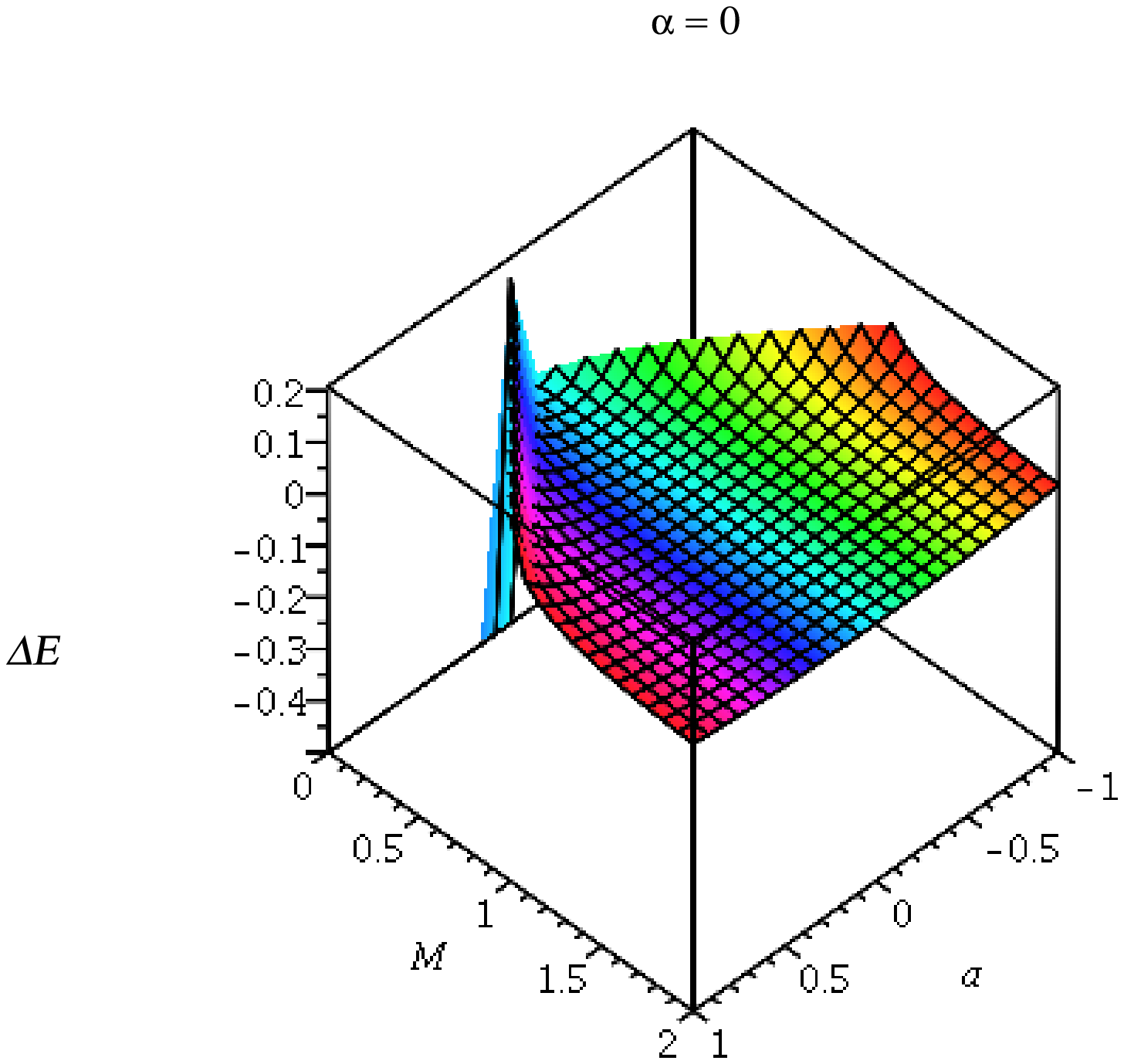}}
{\includegraphics[width=0.45\textwidth]{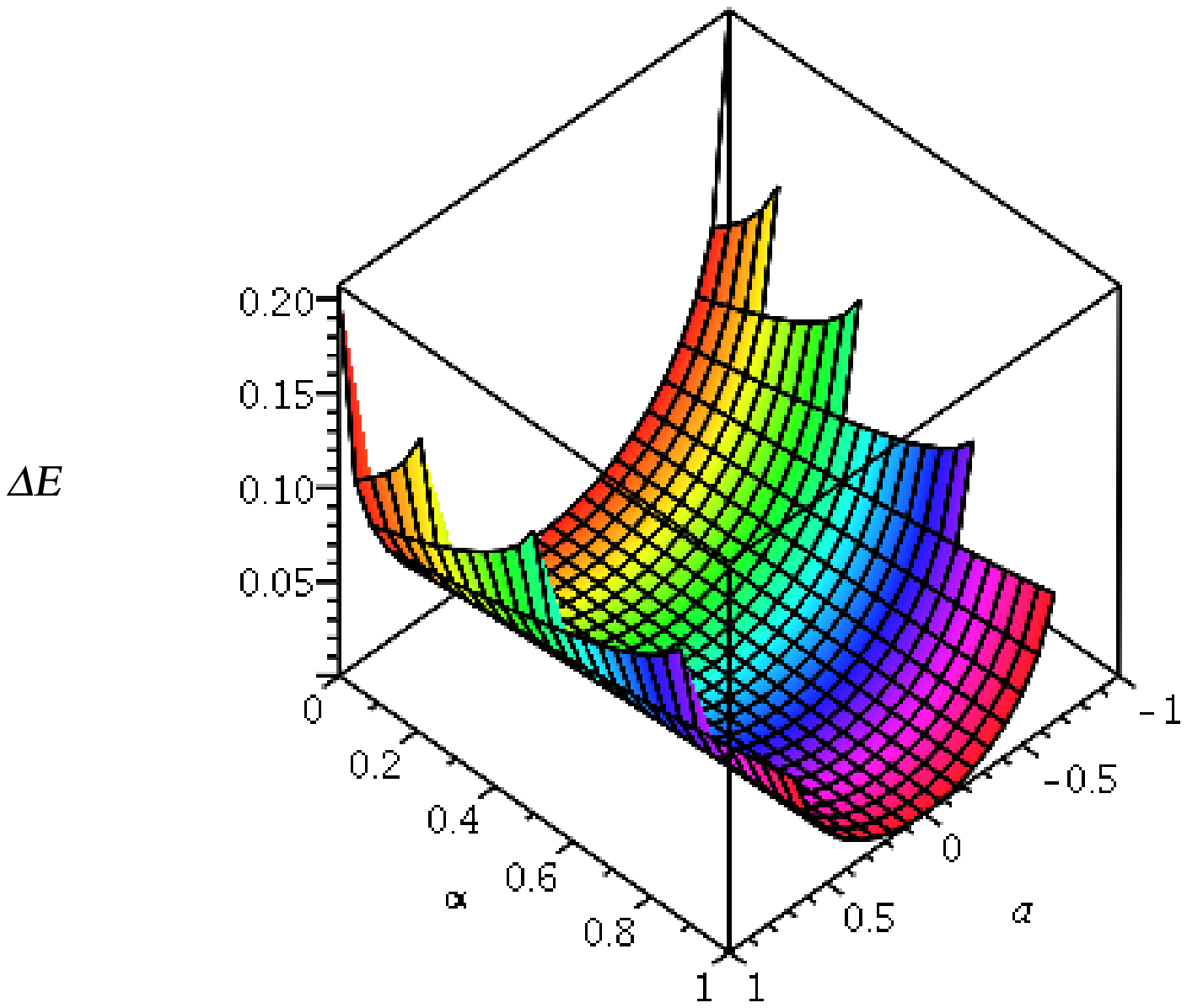}}
{\includegraphics[width=0.45\textwidth]{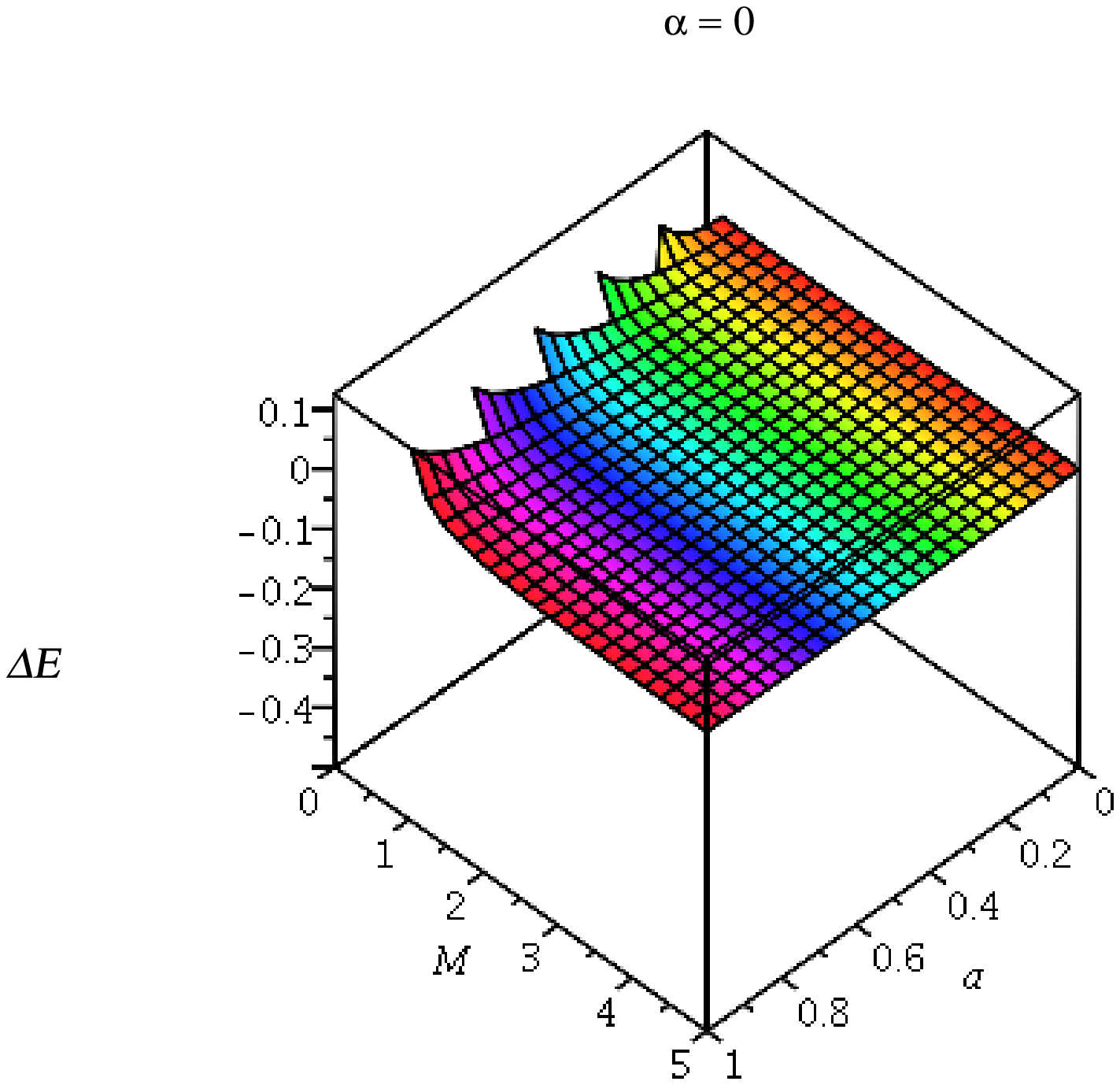}}
{\includegraphics[width=0.45\textwidth]{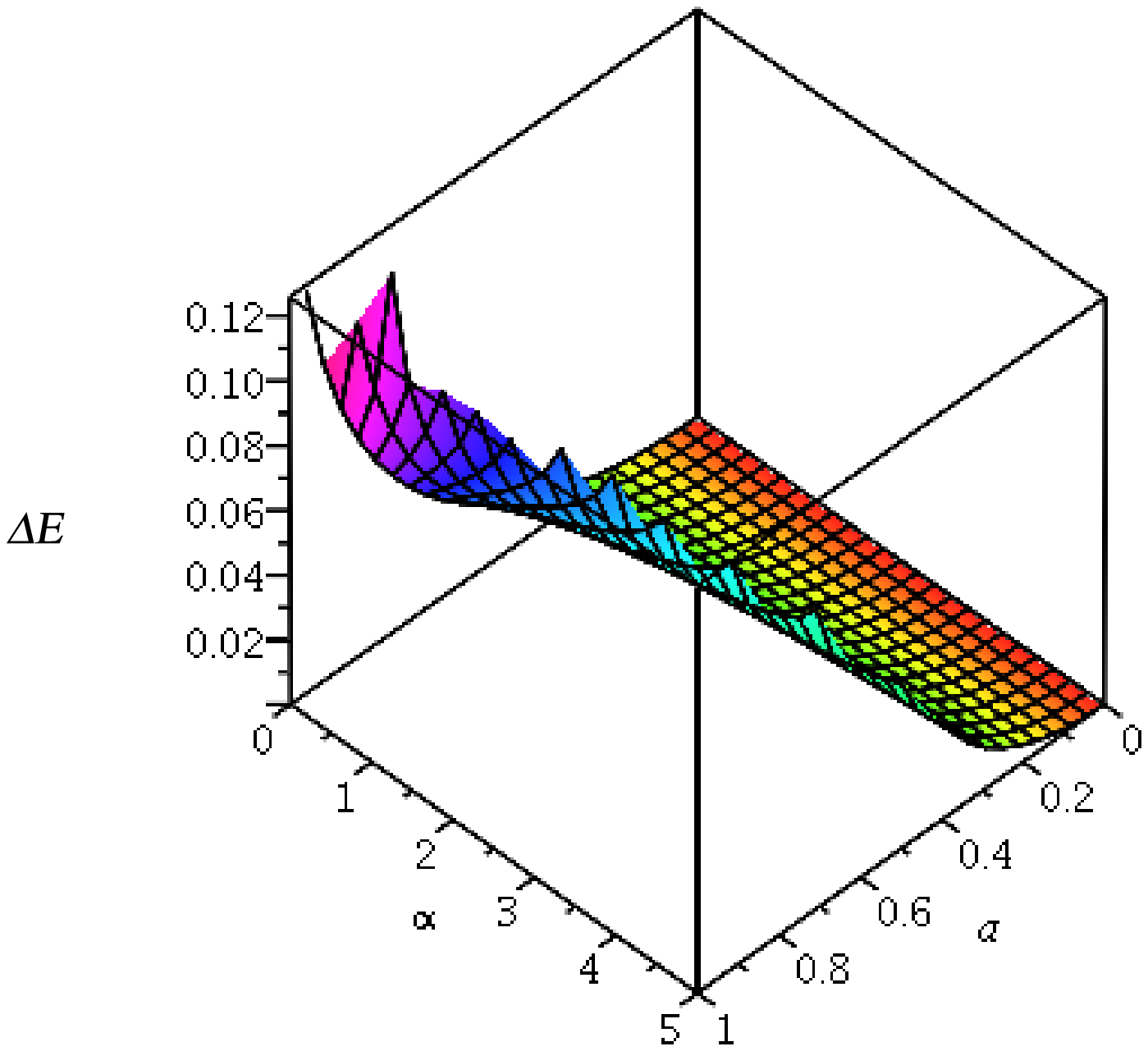}}
\end{center}
\caption{The figure depicts the variation  of $\Delta {\cal E}$  with $a$ and $\alpha$ 
for Kerr BH and Kerr-MOG BH. We have set ${\cal M}=1$.
\label{vfe}}
\end{figure}

The energy gain $\Delta {\cal E}$ in this process becomes 
\begin{eqnarray}
\Delta {\cal E} &=& \frac{1}{2}\left(\frac{\sqrt{2{\cal M}r-\frac{\alpha}{1+\alpha}{\cal M}^2}}{r}-1 \right)
=-{\cal E}^{(x)} ~.\label{mch}
\end{eqnarray}
The maximum gain in energy occurs at the event horizon and this value is given by   
\begin{eqnarray}
\Delta {\cal E} \leq \frac{1}{2}\left(\sqrt{\frac{2{\cal M}}{r_{+}}-\frac{\alpha}{1+\alpha}
\frac{{\cal M}^2}{r_{+}^2} }-1\right) ~.\label{mch1}
\end{eqnarray}
The variation of $\Delta {\cal E}$ with $r_{+}$ could be observed from the Fig.~(\ref{vmf}). 
The gain in energy in terms of spin parameter and MOG parameter is 
\begin{eqnarray}
\Delta {\cal E} \leq \frac{1}{2}\left(\sqrt{\frac{2}{1+\sqrt{\frac{1}{1+\alpha}-\left(\frac{a}{{\cal M}}\right)^2}}
-\frac{\alpha}{1+\alpha} \frac{1}{\left(1+\sqrt{\frac{1}{1+\alpha}-\left(\frac{a}{{\cal M}}\right)^2} \right)^2}}-1\right)
~.\label{mch2}
\end{eqnarray}
This is the \emph{key prediction} of this work.
It is clearly evident that the gain in energy strictly depends upon the MOG parameter. The effect of this parameter 
could be seen from the energy gain versus spin diagram~(Fig.~\ref{vfe1}). From this diagram, one could say that 
there is a direct influence of the MOG parameter in the energy extraction process. When $\alpha=0$, the energy 
gain in Penrose process increases while the spin parameter increases. This scenario is quite different when we 
add the parameter $\alpha$. In this case the energy gain is very slower than the former case. In-fact, the 
energy gain is one half of the previous value.  
When $\alpha=0$, one finds the energy value for Kerr BH. For extremal Kerr-MOG BH, the maximum gain in 
energy is given by 
\begin{eqnarray}
\Delta {\cal E} \leq \frac{1}{2} \left(\sqrt{\frac{2+\alpha}{1+\alpha}}-1 \right)
\end{eqnarray}
It implies that the deformation parameter plays an important role in the energy extraction process, it is 
in fact decreasing the value of $\Delta {\cal E}$ in comparison with extremal Kerr BH. In Fig.~\ref{vfe}, we 
have plotted 3D diagram of energy gain in Penrose process for various parameter space. From these figures 
we can easily see that how the deformation parameter affects in the energy extraction process for Kerr-MOG BH.
\subsection{\label{pp} The Wald Inequality}
It is very important to investigate what is the energy limits in the Penrose process for Kerr-MOG BH? 
In this section, we would try to resolve this issue. Wald~\cite{wald73} was first able to derive 
this limits. He also derived an inequality which explains the origin and the limitations of this 
process. To do this let us consider a particle, with a four velocity $U^{\mu}$ and specific energy 
${\cal E}$, breaks up into fragments. Let $\varepsilon$ be the specific energy and $u^{\mu}$ be the 
four-velocity of one of the fragments. Now we want to derive the limits on $\varepsilon$. 

Choose an orthonormal tetrad-frame, $e_{b}^{\mu}$, in which $U^{\mu}$ coincides with 
$e_{0}^{\mu}$ and the remaining spacelike basis vectors are $e_{(\zeta)}^{\mu}$~($\zeta=1,2,3$):
\begin{eqnarray}
e_{0}^{\mu} =U^{\mu} \,\, \mbox{and}\,\, e_{(\zeta)}^{\mu} 
\end{eqnarray}
In this frame 
\begin{eqnarray}
u^{\mu} &=& \eta \left(U^{\mu}+v^{(\zeta)}e_{(\zeta)}^{\mu}\right) ~.\label{mgu}
\end{eqnarray}
where $v^{(\zeta)}$ are the spatial components of the three-velocity of the fragment 
$\eta=\frac{1}{\sqrt{1-|v|^2}}$ and $|v|^2=v^{(\zeta)}v_{(\zeta)}$. Since the spacetime 
has time-like Killing vector $\xi=\partial_{0}$ then it could be represent in 
tetrad-frame as
\begin{eqnarray}
\xi_{\mu} &=& \xi_{(0)}U_{\mu}+\xi_{(\zeta)}e_{\mu}^{(\zeta)}
\end{eqnarray}
Now the conserved quantity energy ${\cal E}$ could be represent in terms of Killing vector as 
\begin{eqnarray}
{\cal E} &=& -\xi_{\mu}U^{\mu}=-\xi_{(0)}=-\xi^{\mu}U_{\mu}=-\xi^{(0)}, 
\end{eqnarray}
and 
\begin{eqnarray}
g_{00} &=&  \xi_{\mu}\xi^{\mu}=-\xi_{(0)}^2+\xi_{(\mu)}\xi^{(\mu)}=-{\cal E}^2+|\xi|^2 ~.\label{mgu1}.
\end{eqnarray}
Therefore one obtains 
\begin{eqnarray}
|\xi|^2 &=& \xi_{(\mu)}\xi^{(\mu)}= {\cal E}^2+g_{00} ~.\label{mgu2}.
\end{eqnarray}
Using Eq.~(\ref{mgu}), one could obtain the specific energy of the fragment as
\begin{eqnarray}
\varepsilon &=& -\xi_{\mu}u^{\mu}
=\eta \left(\xi_{(0)}+v^{(\zeta)}\xi_{(\zeta)}\right)= \eta \left({\cal E}+|v||\xi|\cos{\vartheta} \right),
~.\label{mgu3}
\end{eqnarray}
where $\vartheta$ is the angle between the three-dimensional vectors $v^{(\zeta)}$ and $\xi_{\mu}$.
Using Eq.~(\ref{mgu1}) and  Eq.~(\ref{mgu2}), one could write the Eq.~(\ref{mgu3}) as
\begin{eqnarray}
\varepsilon &=&  \eta{\cal E}+ \eta |v| \sqrt{{\cal E}^2+g_{00}} \cos{\vartheta} 
~.\label{mgu4}
\end{eqnarray}
This equation provides the inequality 
\begin{eqnarray}
\eta{\cal E}-\eta |v| \sqrt{{\cal E}^2+g_{00}} \le \varepsilon \le 
\eta{\cal E}+ \eta |v| \sqrt{{\cal E}^2+g_{00}}  
~.\label{mmgu5}
\end{eqnarray}
This is called the famous Wald inequality. For Kerr-MOG BH this inequality becomes 
\begin{eqnarray}
\eta{\cal E}-\eta |v| \sqrt{{\cal E}^2+1-\frac{\alpha}{1+\alpha}} \le \varepsilon \le 
\eta{\cal E}+ \eta |v| \sqrt{{\cal E}^2+1-\frac{\alpha}{1+\alpha}} 
~.\label{mgu5}
\end{eqnarray}
We proved that the maximum energy that a particle describing a stable circular 
orbit~(See Appendix: Eq.(\ref{ap6})) is 
\begin{eqnarray}
{\cal E}_{m} &=& \frac{1}{\sqrt{3-\alpha}} ~.\label{mgu6}
\end{eqnarray}
For $\varepsilon$ to be negative, it is thus necessary that 
\begin{eqnarray}
|v| > \frac{{\cal E}}{\sqrt{{\cal E}^2+1-\frac{\alpha}{1+\alpha}}} &=& \frac{\sqrt{1+\alpha}}{2}
~.\label{mgu6.1}
\end{eqnarray}
Otherwise, the fragments must have relativistic energies which becomes possible before any extraction 
of energy by the above process.   

\subsection{\label{pp1} The Bardeen-Press-Teukolsky Inequality}
In this section, we shall review what is the lower bound on the magnitude of three velocity  between 
two particles of different specific energies followed by two orbits and collide at some point~\cite{bpt}. 
Let the two particles have specific energies as ${\cal E}_{1}$ and ${\cal E}_{2}$. Also let the 
magnitude of three velocity between two particles be $\varpi$.

Suppose we have an orthonormal tetrad-frame as defined previously,
\begin{eqnarray}
e_{0}^{\mu} =U^{\mu} \,\, \mbox{and}\,\, e_{(\zeta)}^{\mu}~~~(\zeta=1,2,3),
\end{eqnarray}
in which the two orbits cross with equal and opposite three velocities, +$v^{(\zeta)}$ 
and -$v^{(\zeta)}$ so that 
\begin{eqnarray}
\varpi &=& \frac{2|v|}{1+|v|^2}\,\, \mbox{where} \,\, |v|^2=v^{(\zeta)}v_{(\zeta)} ~.\label{mgw}
\end{eqnarray}
The four velocities, $u_{1}^{\mu}$ and $u_{2}^{\mu}$  of two particles in the said 
tetrad-frame at the time of collision are 
\begin{eqnarray}
u_{1}^{\mu} &=& \eta \left(U^{\mu}+v^{(\zeta)}e_{(\zeta)}^{\mu}\right) ~.\label{mgu7}\\
u_{2}^{\mu} &=& \eta \left(U^{\mu}-v^{(\zeta)}e_{(\zeta)}^{\mu}\right) ~.\label{mgu8}
\end{eqnarray}
where $\eta=\frac{1}{\sqrt{1-|v|^2}}$.
As proceeding previously the space-time allows a time-like Killing vector 
$\xi=\partial_{0}$ then its representation in tetrad-frame be
\begin{eqnarray}
\xi^{\mu} &=& \xi^{(0)}U^{\mu}+\xi^{(\zeta)}e_{(\zeta)}^{\mu}\\
\xi_{\mu} &=& \xi_{(0)}U_{\mu}+\xi_{(\chi)}e_{\mu}^{(\chi)}\,\, (\xi^{(0)}=\xi_{(0)})
\end{eqnarray}
Now, by definition,
\begin{eqnarray}
g_{00} &=&- \xi^{\mu}\xi_{\mu}=-\xi^{(0)}\xi_{(0)}+\xi^{(\zeta)}\xi_{(\zeta)}=
-\xi_{(0)}^2+|\xi|^2~, \label{mgu9}.
\end{eqnarray}
so that 
\begin{eqnarray}
|\xi|^2=\xi_{(0)}^2+g_{00}~.\label{mgu10}
\end{eqnarray}
The specific energies at the time of collision are given by 
\begin{eqnarray}
{\cal E}_{1} &=&- \xi_{\mu}u^{\mu}
=\eta \left(\xi_{(0)}+v^{(\zeta)}\xi_{(\zeta)}\right)= \eta \left(\xi_{(0)}+|v||\xi|\cos{\vartheta} \right),
~\label{mgu11}
\end{eqnarray}
and 
\begin{eqnarray}
{\cal E}_{2}  &=& -\xi_{\mu}u^{\mu}
=\eta \left(\xi_{(0)}-v^{(\zeta)}\xi_{(\zeta)}\right)= \eta \left(\xi_{(0)}-|v||\xi|\cos{\vartheta} \right)
~.\label{mgu12}
\end{eqnarray}
where $\vartheta$ is the angle between the 3-vectors $v^{(\zeta)}$ and $\xi_{\mu}$. From the preceeding equations 
we can write 
\begin{eqnarray}
{\cal E}_{1}+{\cal E}_{2}  &=& 2\eta \xi_{(0)} ~.\label{mgu13}\\
{\cal E}_{1}-{\cal E}_{2}  &=& 2\eta |v||\xi|\cos{\vartheta}
\end{eqnarray}
Therefore, 
\begin{eqnarray}
\left({\cal E}_{1}-{\cal E}_{2}\right)^2  &=& 4\eta^2 |v|^2|\xi|^2\cos^2{\vartheta}\\
&=& |v|^2 \left(4\eta^2 \xi_{(0)}^2+4\eta^2g_{00} \right) \cos^2{\vartheta} \\
&=& |v|^2 \left[\left({\cal E}_{1}+{\cal E}_{2}\right)^2 +4\eta^2 g_{00}  \right]\cos^2{\vartheta}
\end{eqnarray}
It indicates that 
\begin{eqnarray}
\left({\cal E}_{1}-{\cal E}_{2}\right)^2  
& \leq & |v|^2 \left[\left({\cal E}_{1}+{\cal E}_{2}\right)^2 +4\eta^2 g_{00} \right];
\end{eqnarray}
Substituting the value of $\eta$, one obtains 
\begin{eqnarray}
|v|^2 \left[\left({\cal E}_{1}+{\cal E}_{2}\right)^2 +\frac{4}{1-|v|^2} g_{00} \right] & \geq &
\left({\cal E}_{1}-{\cal E}_{2}\right)^2,
\end{eqnarray}
or re-arranging this equation 
\begin{eqnarray}
-|v|^4 \left({\cal E}_{1}+{\cal E}_{2}\right)^2+2|v|^2\left({\cal E}_{1}^2+{\cal E}_{2}^2+2g_{00} \right)
-\left({\cal E}_{1}-{\cal E}_{2}\right)^2 & \geq & 0
\end{eqnarray}
It follows that 
\begin{eqnarray}
|v| & \geq & \frac{\left|\sqrt{{\cal E}_{1}^2+g_{00}}-\sqrt{{\cal E}_{2}^2+g_{00}}\right|}{{\cal E}_{1}+{\cal E}_{2}} 
~.\label{mgu15}
\end{eqnarray}
and the required lower bound on $\varpi$ according to Eq.~(\ref{mgw}); and consequently the inequality 
is called well-known Bardeen-Press-Teukolsky inequality~\cite{bpt}.

In case of Kerr-MOG BH, let the particle with the energy ${\cal E}_{1}$ followed by a stable circular 
geodesics in the equatorial plane then its maximum energy is given in Eq.~(\ref{mgu6}). Since the value 
of $g_{00}=1-\frac{\alpha}{1+\alpha}$ and choosing the value of ${\cal E}_{2}=0$, the 
inequality~(\ref{mgu15}) becomes 
\begin{eqnarray}
|v| & > & \frac{2-\sqrt{3-\alpha}}{\sqrt{1+\alpha}}     ~.\label{mgu16}
\end{eqnarray}
and subsequently the inequality for $\varpi$ is 
\begin{eqnarray}
\varpi &\geq& \frac{\sqrt{1+\alpha}}{2} ~.\label{mgu17}
\end{eqnarray}
which is in agreement with the result~(\ref{mgu6.1}) performed from Wald's inequality. In the limit 
$\alpha=0$, one gets the result for Kerr BH. The key conclusion from the two ineqalities are that 
to achieve effective energy extraction from Penrose process,  one should first accelerate the particle 
pieces to more than $\frac{\sqrt{1+\alpha}}{2}$ times the speed of light by hydrodynamical forces.

\subsection{\label{ree} The Irreducible Mass \&  Reversible Extraction of Energy}
In a landmark paper ``Reversible Transformations of a Charged Black Hole''~\cite{cr71}, Christodoulou 
and Ruffini have derived an important relation between energy of a charged rotating BH and the irreducible 
mass~\cite{cd70} of the BH. Using similar analogy, in this section we would like to provide the relation between 
the energy and the irreducible mass for Kerr-MOG BH. It is now well established by fact that the 
BH area never decreases.

To prove the area of the BH always increases, we could define the ``irreducible mass''~\cite{sean} as
\begin{eqnarray}
{\cal M}_{irr} &=& \sqrt{\frac{{\cal A}}{16\pi G^2}} ~. \label{mi}
\end{eqnarray}
For Kerr-MOG BH, it is given by 
\begin{eqnarray}
{\cal M}_{irr} &=& \frac{\sqrt{\left(\frac{\alpha+2}{\alpha+1}\right){\cal M}^2+2 
\sqrt{\frac{{\cal M}^4}{1+\alpha} -J^2}}}{2(1+\alpha)} ~. \label{mi1}
\end{eqnarray}
Using this definition, the inequality~(\ref{mch1}) becomes
\begin{eqnarray}
\Delta {\cal E} \leq \frac{1}{2} \left[\frac{{\cal M}}{{\cal M}_{irr}}
\left\{\left(1+\alpha \right)^2+\left(\frac{\alpha}{1+\alpha}\right)
\frac{{\cal M}^2}{4 {\cal M}_{irr}^2} \right\}^{-1}-1\right] ~.\label{mi2}
\end{eqnarray}
One could derive more general inequality by using Eq.~(\ref{mengpp}) if and only if 
\begin{eqnarray}
\left[r^4+a^2r^2+2{\cal M}a^2r-\frac{\alpha}{1+\alpha}{\cal M}^2a^2 \right] {\cal E} -
a\left(2{\cal M}r-\frac{\alpha}{1+\alpha}{\cal M}^2 \right) \ell \geq 0 ~.\label{mi3}
\end{eqnarray}
The inequality should be equality if the process considered occurs at the outer horizon i.e.
\begin{eqnarray}
\left[(r_{+}^2+a^2)r_{+}^2+2{\cal M}a^2r_{+}-\frac{\alpha}{1+\alpha}{\cal M}^2a^2\right] {\cal E} -
a\left(2{\cal M}r_{+}-\frac{\alpha}{1+\alpha}{\cal M}^2 \right) \ell \geq 0 ~.\label{mi4}
\end{eqnarray}
Let a particle with negative energy, $-{\cal E}$ and an angular momentum, $-{\ell}$ approaching towards 
the outer horizon then the gain in energy $\delta {\cal M}(={\cal E})$ and the gain in the angular momentum 
$\delta J(=\ell)$ under the condition
\begin{eqnarray}
\left[(r_{+}^2+a^2)r_{+}^2+2{\cal M}a^2r_{+}-\frac{\alpha}{1+\alpha}{\cal M}^2a^2\right] \delta {\cal M} \geq
a\left(2{\cal M}r_{+}-\frac{\alpha}{1+\alpha}{\cal M}^2 \right) \delta J ~.\label{mi5}
\end{eqnarray}
Let us consider the process should take place adiabatically then 
\begin{eqnarray}
\delta J= {\cal M} \delta a+a\delta {\cal M}
\end{eqnarray}
Therefore the inequality~(\ref{mi5}) becomes 
\begin{eqnarray}
\left(r_{+}^2+a^2 \right)r_{+}^2 \delta {\cal M} \geq 
a{\cal M}\left(2{\cal M}r_{+}-\frac{\alpha}{1+\alpha}{\cal M}^2 \right) \delta a ~.\label{mi6}
\end{eqnarray}
More precisely, this can be written as 
\begin{eqnarray}
r_{+}^2 \delta {\cal M} \geq a{\cal M} \delta a ~.\label{mi7}
\end{eqnarray}
By the definition of irreducible mass it has been shown that for Kerr BH
\begin{eqnarray}
\delta {\cal M}_{irr}^2 &=& \frac{r_{+}^2 \delta {\cal M} - a{\cal M} \delta a}{2\sqrt{{\cal M}^2-a^2}}  
~. \label{mi8}
\end{eqnarray}
Using same analogy, one could say that for Kerr-MOG BH
\begin{eqnarray}
\delta {\cal M}_{irr}^2 \geq 0
\end{eqnarray}
It implies that by no continuous process it is impossible to decrease the irreducible mass of a BH. We can 
also say that by no continuous process it is impossible to decrease the surface area of a BH. Where the 
surface area of a BH can be defined as 
\begin{eqnarray}
{\cal A}=4\pi \left(r_{+}^2+a^2 \right) =16\pi G^2 {\cal M}_{irr}^2.~\label{ar}
\end{eqnarray}
We determine the rotational energy as 
\begin{eqnarray}
{\cal E}_{Rot}={\cal M}-{\cal M}_{irr}={\cal M}-\frac{1}{2(1+\alpha)}
\left[\left(\frac{\alpha+2}{\alpha+1}\right){\cal M}^2+\sqrt{\frac{4{\cal M}^4}{1+\alpha}-4J^2} \right]^{\frac{1}{2}}
~. \label{mi9}
\end{eqnarray}
For higher dimensional BH and black ring this has been studied by Nozawa et al.~\cite{maeda}.

\begin{figure}
\begin{center}
{\includegraphics[width=0.45\textwidth]{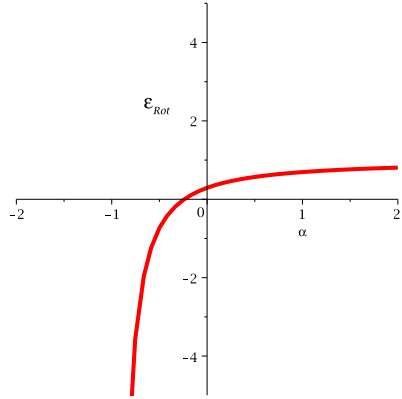}}
\end{center}
\caption{The figure depicts the variation  of $\varepsilon_{Rot}$  with $\alpha$.
\label{vfer}}
\end{figure}

For extremal Kerr-MOG BH, one gets the ratio as 
\begin{eqnarray}
\varepsilon_{Rot}=\frac{{\cal E}_{Rot}}{{\cal M}}
=1-\frac{1}{2(1+\alpha)}\sqrt{\frac{\alpha+2}{\alpha+1}}
~. \label{mi10}
\end{eqnarray}
when $\alpha=0$, $\varepsilon_{Rot}\simeq 29$ percentage. When $\alpha\neq 0$, $\varepsilon_{Rot}$ varies 
as in the  Fig.~\ref{vfer}.
Using Eq.~(\ref{ar}), one could say that by ``no continuous process can the surface area of a BH be decreased''
~\cite{sch}.  This is the outcome of Hawking's area theorem. It should be emphasized that the irreducible mass of 
a BH never be unchanged and the processes in which it should remain constant are said to be reversible one. We also 
should noted that by virtue of definition~(\ref{mi1}), the Christodoulou-Ruffini mass formula for Kerr-MOG BH 
becomes 
\begin{eqnarray}
 {\cal M}^2 &=& \left[(1+\alpha) {\cal M}_{irr}+\frac{\alpha}{(1+\alpha)^2} \frac{{\cal M}^2}{4 {\cal M}_{irr}}\right]^2
 +\frac{J^2}{4(1+\alpha)^2 {\cal M}_{irr}^2}
\end{eqnarray}
Now let us pause! What is the physical meaning of this equation. It indicates that if ${\cal M}_{irr}$ is 
irreducible one then the second term $\frac{J^2}{4(1+\alpha)^2 {\cal M}_{irr}^2}$ gives us towards the 
contribution of the rotational kinetic energy to the square of the inertial mass of the BH. This means 
that it is the rotational energy which is being extracted by the Penrose process. 

\section{Epicyclic Frequencies in  Kerr-MOG BH}
In this section, we shall review the orbital epicyclic frequencies which could be derived from the effective 
potential for circular geodesics in MOG. The derivation of this frequencies could be directly computed from 
the concept of conservation of energy and conservation of angular momentum. The effective potential concept 
also help us to compute these frequencies. Finally, we have discussed the astropphysical applications of these 
frequencies i.e. the QPO. QPOs are a common feature of X-ray flux of steller mass BHs. To get the appropriate 
information on the spacetime geometry around the stellar mass BH, QPOs are very useful tool. Aspects of circular 
geodesic properties have been studied for various class of BHs in  many years due to the fundamental role in 
accretion-disk physics. The said circular geodesics could be expressed in terms of three fundamental frequencies: the 
Keplerian frequency, the radial and  vertical epicyclic frequencies. It must be noted that these frequencies 
are depend on structure of the geometry of the space-time.  These frequencies are also function of mass parameter, 
radial parameter and spin parameter. 

In Newton's gravity, these three characteristic frequencies are 
same when the potential as $\Phi=-\frac{M}{r}$ i.e. 
\begin{eqnarray}
\nu_{\phi}=\nu_{\theta}=\nu_{r}= \frac{M}{r^{\frac{3}{2}}}
\end{eqnarray}
The equality of these three frequencies indicate that the orbits in the $\Phi=-\frac{M}{r}$ are periodic 
and closed. In order to derive the fundamental frequencies in Kerr-MOG spacetime we have to consider 
the general stationary and axisymmetric spacetime as follows
\begin{eqnarray}
 ds^2=g_{tt}dt^2+g_{rr}dr^2+g_{\theta\theta}d\theta^2+g_{\phi\phi}d\phi^2+2g_{t\phi}d\phi dt,
 \label{km1}
\end{eqnarray}
where $g_{\mu\nu}=g_{\mu\nu}(r, \theta)$. It follows that the metric components are independent 
of the time ~$t$  and  $\phi$ coordinates. It immediately suggests that there exists
two constants of motion: the conserved specific energy ${\cal E}$ and the conserved specific 
angular momentum~$\ell$. Thus the four-velocity components of $t$ and $\phi$ are 
\begin{eqnarray}
\dot{\phi} &=& -\frac{ g_{t\phi} {\cal E}+g_{tt} \ell}{g_{t\phi}^2-g_{tt}g_{\phi\phi}}~.\label{phi1}\\
\dot{t} &=& \frac{g_{\phi\phi} {\cal E}+g_{t\phi} \ell}{g_{t\phi}^2-g_{tt}g_{\phi\phi}}  ~.\label{mut}
\end{eqnarray}
From the normalization condition of four velocity $g_{\mu\nu} u^{\mu}u^{\nu}=-1$, we get 
\begin{eqnarray}
g_{rr} \dot{r}^2+g_{\theta\theta} \dot{\theta}^2 &=& {\cal V}_{eff}~(r,\theta,{\cal E}, \ell)~.\label{mut1}
\end{eqnarray}
Therefore the effective potential could be defined as 
\begin{eqnarray}
{\cal V}_{eff} &=& \frac{({\cal E}^2+g_{tt})g_{\phi\phi}+(2\ell{\cal E}-g_{t\phi})g_{t\phi}+\ell^2g_{tt}}
{g_{t\phi}^2-g_{tt}g_{\phi\phi}}  ~.\label{vef}
\end{eqnarray}
For circular orbits in the equatorial plane one has $\dot{r}=\dot{\theta}=0$, which directly implies 
${\cal V}_{eff}=0$, and $\ddot{r}=\ddot{\theta}=0$ which gives $\partial_{r}{\cal V}_{eff}=0$ and 
$\partial_{\theta}{\cal V}_{eff}=0$ respectively. From these conditions one can obtain the energy and 
angular momentum~\cite{bambi} as 
\begin{eqnarray}
 {\cal E} &=& -\frac{g_{tt}+\Omega_{\phi} g_{t\phi}}{\sqrt{-g_{tt}-2g_{t\phi}\Omega_{\phi}-g_{\phi\phi}\Omega_{\phi}^2}}
\end{eqnarray}
and 
\begin{eqnarray}
\ell &=& \frac{g_{t\phi}+\Omega_{\phi} g_{\phi\phi}}{\sqrt{-g_{tt}-2g_{t\phi}\Omega_{\phi}-g_{\phi\phi}\Omega_{\phi}^2}}
\end{eqnarray}
Now the proper angular momentum ~($l$) of a test particle can be 
derived as 
\begin{eqnarray}
 l=-\frac{g_{t\phi}+\Omega_{\phi} g_{\phi\phi}}{g_{tt}+\Omega_{\phi} g_{t\phi}},
\end{eqnarray}
where, $\Omega_{\phi}$ is the orbital frequency of a test particle. Now the $\Omega_{\phi}$
can be defined as
\begin{eqnarray}
\Omega_{\phi} \equiv  2\pi\nu_{\phi} =\frac{\dot{\phi}}{\dot{t}}=\frac{(\frac{d\phi}{d\tau})}{(\frac{dt}{d\tau})} 
=\frac{d\phi}{dt}
=\frac{-\partial_{r}g_{t\phi}\pm \sqrt{(\partial_{r}g_{t\phi})^2-(\partial_{r}g_{tt})(\partial_{r}g_{\phi\phi})}}
{\partial_{r}g_{\phi\phi}} 
 \label{km2}
\end{eqnarray}
The upper sign is for corotating orbit and the lower sign is for counterrotating orbit. If 
$\partial_{r}^2{\cal V}_{eff} \le 0$ and $\partial_{\theta}^2{\cal V}_{eff} \le 0$ then the 
orbits are stable under small perturbations.

For Kerr-MOG BH, the Kepler frequency is derived to be 
\begin{eqnarray}
\Omega_{\phi}^{d}=\frac{\sqrt{G_{N}{\cal M}r-\frac{\alpha}{1+\alpha} {\cal M}^2}}
 {r^2+a \sqrt{G_{N}{\cal M}r-\frac{\alpha}{1+\alpha} {\cal M}^2}} \label{km3}
\end{eqnarray}
and
\begin{eqnarray}
\Omega_{\phi}^{g}=-\frac{\sqrt{G_{N}{\cal M}r-\frac{\alpha}{1+\alpha} {\cal M}^2}}
 {r^2-a \sqrt{G_{N}{\cal M}r-\frac{\alpha}{1+\alpha} {\cal M}^2}} ~\label{km4}.
\end{eqnarray}
where the negative sign implies that the rotation is in the reverse direction. Suffixes $d$ and
$g$ denote for the direct orbit and retrograde orbit respectively.

The general expressions for computing the radial~($\Omega_{r}$) and vertical
($\Omega_{\theta}$) epicyclic frequencies are~\cite{doneva,jcapcpp}
\begin{eqnarray}\nonumber
\Omega_{r}^2 &=& \frac{(g_{tt}+\Omega_{\phi}g_{t\phi})^2}{2~g_{rr}}~\partial_{r}^2~U\\
 &=& \frac{(g_{tt}+\Omega_{\phi}g_{t\phi})^2}{2~g_{rr}}
\left[\partial_{r}^2\left(\frac{g_{\phi\phi}}{Y}\right) +2l~\partial_{r}^2\left(\frac{g_{t\phi}}{Y}\right)
+l^2~\partial_{r}^2\left(\frac{g_{tt}}{Y}\right)\right]|_{r=const.,~\theta=\frac{\pi}{2}}\nonumber \label{km5}
\end{eqnarray}
and
\begin{eqnarray}\nonumber
\Omega_{\theta}^2 &=& \frac{(g_{tt}+\Omega_{\phi}g_{t\phi})^2}{2~g_{\theta\theta}}~\partial_{\theta}^2~U\\
 &=& \frac{(g_{tt}+\Omega_{\phi}g_{t\phi})^2}{2~g_{\theta\theta}}\left[\partial_{\theta}^2\left(\frac{g_{\phi\phi}}{Y}\right)
 +2l~\partial_{\theta}^2\left(\frac{g_{t\phi}}{Y}\right)+l^2~\partial_{\theta}^2\left(\frac{g_{tt}}{Y}\right)\right]|_{r=const.
 ~,\theta=\frac{\pi}{2}}\nonumber \label{km6}
\end{eqnarray}
respectively and $Y$ can be defined as
\begin{eqnarray}
 Y=g_{tt}g_{\phi\phi}-g_{t\phi}^2 .
\end{eqnarray}
The conditions $\Omega_{r}^2\geq 0$ and $\Omega_{\theta}^2\geq 0$ implies that stability of 
the circular geodesic motions against small oscillations. From the condition of radial 
stability one can determined the radii of ISCO. For example, it is well known that the ISCO is 
located for Schwarzschild BH at $r=r_{isco}=6M$ while for extremal Kerr BH the ISCO is located at 
$r_{isco}=M$ for direct orbit and $r_{isco}=9M$ for retrograde orbit~\cite{sch}. It should be 
noted that for non-negative value of $\Omega_{\theta}$ indicates that the geodesic motion is 
stable under small oscillations in the vertical direction.

Since we are restricted in the equatorial plane thus $\theta=\frac{\pi}{2}$. The proper angular 
momentum for the equatorial plane is calculated to be 
\begin{eqnarray}
 l^{d} &=& \frac{(r^2+a^2)\sqrt{G_{N}{\cal M}r-\frac{\alpha}{1+\alpha} {\cal M}^2}-2aG_{N}{\cal M}r
 +\frac{\alpha}{1+\alpha}a {\cal M}^2 }{r^2-2G_{N}{\cal M}r+\frac{\alpha}{1+\alpha}{\cal M}^2 
 +a\sqrt{G_{N}{\cal M}r-\frac{\alpha}{1+\alpha} {\cal M}^2}} 
 \label{km7}
\end{eqnarray}
and
\begin{eqnarray}
 l^{g} &=&- \frac{(r^2+a^2)\sqrt{G_{N}{\cal M}r-\frac{\alpha}{1+\alpha} {\cal M}^2}+2aG_{N}{\cal M}r
 -\frac{\alpha}{1+\alpha}a {\cal M}^2 }{r^2-2G_{N}{\cal M}r+\frac{\alpha}{1+\alpha}{\cal M}^2 
 -a\sqrt{G_{N}{\cal M}r-\frac{\alpha}{1+\alpha} {\cal M}^2}} 
 \label{km8}
\end{eqnarray}
It should be noted that for Kerr-MOG BH, $Y=-\Delta$ and 
\begin{eqnarray}
 \partial_{r}^2~U &=& \frac{2 {\cal F}(r)}
 {\Delta \left(r^2-2G_{N}{\cal M}r+\frac{\alpha}{1+\alpha}{\cal M}^2
 \pm a\sqrt{G_{N}{\cal M}r-\frac{\alpha}{1+\alpha} {\cal M}^2} \right)^2}
\end{eqnarray}
where
$$
{\cal F}(r)=G_{N}{\cal M}r^3-6G_{N}^2{\cal M}^2r^2 
 +9\frac{\alpha}{1+\alpha} G_{N}{\cal M}^3r-3G_{N}{\cal M}a^2 r
$$
$$
 \pm 8a\left(G_{N}{\cal M}r-\frac{\alpha}{1+\alpha} {\cal M}^2\right)^\frac{3}{2}
 +4\frac{\alpha}{1+\alpha}{\cal M}^2\left(a^2-\frac{\alpha}{1+\alpha} {\cal M}^2 \right)
$$
which is calculated at $r=const$ and $\theta=\pi/2$. The upper sign indicates for direct orbit 
and lower sign indicates for retrograde orbit respectively. The value of ${\cal F}(r)$ can 
be rewritten as
$$
{\cal F}(r)=G_{N}{\cal M}r \Delta-4\left(G_{N}{\cal M}r-\frac{\alpha}{1+\alpha} {\cal M}^2\right)
\left[\sqrt{G_{N}{\cal M}r-\frac{\alpha}{1+\alpha} {\cal M}^2} \mp a\right]^2
$$
Therefore, we get the radial epicyclic frequencies $\Omega_{r}$ for the direct rotation and retrograde 
rotation as 
\begin{eqnarray}
 \Omega_{r}^{2~(d)} &=& \frac{G_{N}{\cal M}r \Delta-4\left(G_{N}{\cal M}r-\frac{\alpha}{1+\alpha} {\cal M}^2\right)
\left[\sqrt{G_{N}{\cal M}r-\frac{\alpha}{1+\alpha} {\cal M}^2} - a\right]^2}
{\left(r^2+a \sqrt{G_{N}{\cal M}r-\frac{\alpha}{1+\alpha} {\cal M}^2} \right)^2}
 \label{km9}
\end{eqnarray}
and
\begin{eqnarray}
\Omega_{r}^{2~(g)} &=& \frac{G_{N}{\cal M}r \Delta-4\left(G_{N}{\cal M}r-\frac{\alpha}{1+\alpha} {\cal M}^2\right)
\left[\sqrt{G_{N}{\cal M}r-\frac{\alpha}{1+\alpha} {\cal M}^2} + a\right]^2}
{\left(r^2-a \sqrt{G_{N}{\cal M}r-\frac{\alpha}{1+\alpha} {\cal M}^2} \right)^2}
  \label{km10}
\end{eqnarray}
respectively. Setting $\Omega_{r}^{2}=0$, we obtain the ISCO equation for Kerr-MOG BH. Now we can define the 
periastron precession frequency for direct rotation as 
\begin{eqnarray}
 \Omega_{per}^d &=& \Omega_{\phi}^d-\Omega_{r}^d 
 \label{km11}
\end{eqnarray}
which is calculated to be 
\begin{eqnarray}
\Omega_{per}^d &=& \frac{{\cal G}(-)}
{r\left(r^2+a \sqrt{G_{N}{\cal M}r-\frac{\alpha}{1+\alpha} {\cal M}^2} \right)} 
 \label{km12}
\end{eqnarray}
and for retrograde rotation the precession frequency is
\begin{eqnarray}
\Omega_{per}^{g} &=& \Omega_{\phi}^{g}-\Omega_{r}^{g} \nonumber\\
&=&-\frac{{\cal G}(+)}
{r\left(r^2-a \sqrt{G_{N}{\cal M}r-\frac{\alpha}{1+\alpha} {\cal M}^2} \right)}
\label{km13}
 \end{eqnarray}
where 
$$
{\cal G(\mp)} = r\sqrt{G_{N}{\cal M}r-\frac{\alpha}{1+\alpha} {\cal M}^2}
 \mp 
$$
$$
\sqrt{G_{N}{\cal M}r \Delta-4\left(G_{N}{\cal M}r-\frac{\alpha}{1+\alpha} {\cal M}^2\right)
\left[\sqrt{G_{N}{\cal M}r-\frac{\alpha}{1+\alpha} {\cal M}^2} \mp a\right]^2}
$$

To compute the orbital planer precession frequency first we have to calculate the 
vertical epicyclic frequency and to get it we have to derive 
\begin{eqnarray}
 \partial_{\theta}^2~U &=& \frac{2 {\cal H}(\mp)}
 {\left(r^2-2G_{N}{\cal M}r+\frac{\alpha}{1+\alpha}{\cal M}^2
 \pm a\sqrt{G_{N}{\cal M}r-\frac{\alpha}{1+\alpha} {\cal M}^2} \right)^2}
\end{eqnarray}
where
$$
{\cal H}(\mp)=G_{N}{\cal M}r^3-\frac{\alpha}{1+\alpha}{\cal M}^2r^2 
\mp 2 a \left(2 G_{N}{\cal M} r-\frac{\alpha}{1+\alpha}{\cal M}^2 \right)
\sqrt{G_{N}{\cal M}r-\frac{\alpha}{1+\alpha} {\cal M}^2} 
$$
$$
+ a^2 \left(3 G_{N}{\cal M} r-2 \frac{\alpha}{1+\alpha}{\cal M}^2 \right)
$$
which is evaluated at $r=const$ and $\theta=\pi/2$. The upper~(lower) sign 
indicates for direct~(retrograde) orbit respectively. 

Analogously, we get the vertical epicyclic frequencies $\Omega_{\theta}$ for 
direct rotation and retrograde rotation as 
\begin{eqnarray}
\Omega_{\theta}^{2~(d)} &=& \frac{{\cal H}(-)}
{\left(r^2+a\sqrt{G_{N}{\cal M}r-\frac{\alpha}{1+\alpha} {\cal M}^2} \right)^2}
\label{km14}
\end{eqnarray}
and
\begin{eqnarray}
\Omega_{\theta}^{2~(g)} &=& \frac{{\cal H}(+)}
{\left(r^2-a \sqrt{G_{N}{\cal M}r-\frac{\alpha}{1+\alpha} {\cal M}^2} \right)^2}
  \label{km15}
\end{eqnarray}
respectively.

Now we have the value of Keplerian frequency and vertical epicyclic frequency as 
derived previously therefore we can easily compute the nodal precession frequency. 
It is also said to be orbital planer precession frequency or the Lense-Thirring~(LT) 
precession frequency of a test particle. Thus, we get the nodal precession frequency 
for direct rotation as
\begin{eqnarray}
\Omega_{nod}^d &=& \Omega_{\phi}^d-\Omega_{\theta}^d  \label{km16}
\end{eqnarray}
which is calculated to be 
\begin{eqnarray}
\Omega_{nod}^d &=& \frac{r\sqrt{G_{N}{\cal M}r-\frac{\alpha}{1+\alpha} {\cal M}^2}- \sqrt{{\cal H}(-)}}
{r\left(r^2+a \sqrt{G_{N}{\cal M}r-\frac{\alpha}{1+\alpha} {\cal M}^2} \right)} 
 \label{km17}
\end{eqnarray}
while for retrograde rotation it is 
\begin{eqnarray}
\Omega_{nod}^g &=&- \frac{r\sqrt{G_{N}{\cal M}r-\frac{\alpha}{1+\alpha} {\cal M}^2}+ \sqrt{{\cal H}(+)}}
{r\left(r^2-a \sqrt{G_{N}{\cal M}r-\frac{\alpha}{1+\alpha} {\cal M}^2} \right)} 
 \label{km18}
\end{eqnarray}
Negative sign confirms the rotation is in the reverse direction.

\section{\label{dis} Discussion and Outlook}
The study of this work is two-fold. In first part, we explored on the study of energy 
extraction by the Penrose process for Kerr-MOG BH.  We derived the gain in energy for 
said BH. It is deriven in Eq.~(\ref{mch2}). If $\alpha=0$, one obtains the gain 
in energy  for Kerr BH. For extremal Kerr-MOG BH, we derived the maximum gain in energy 
is $\Delta {\cal E} \leq \frac{1}{2} \left(\sqrt{\frac{\alpha+2}{1+\alpha}}-1 \right)$. 
We showed that the MOG parameter has an important role in the energy extraction process 
and it is in fact reduced the value of $\Delta {\cal E}$ in contrast with extremal 
Kerr BH. Finally, we described the Wald inequality and the Bardeen-Press-Teukolsky 
inequality for Kerr-MOG BH in comparison with Kerr BH.  It would be an interesting 
project if one could study the Blandford-Znajek process~\cite{bz} for this BH 
where one may extract the rotational energy by electromagnetically from spinning BH. 

In scecond part, we studied the strong gravity effect of the geodesic motion in terms of 
three fundamental frequencies: the Keplerian frequency, the radial epicyclic frequency 
and the vertical epicyclic frequency. We derived three characteristic frequencies to examine 
the strong gravity effect near the BH. We used the concept of effective potential method 
and the laws of conservation of energy, and angular momentum. 
The stability analysis has been carried out in the radial and vertical directions by using 
characteristic frequencies. The ISCO condition is derived by using the radial epicyclic frequency.
Unlike in Newtonian gravity where all three characteristic frequencies are equal, we observed 
in modified gravity that these frequencies have different value indicates the strong gravity 
effects near the BHs. Finally, we computed the periastron precession frequency and the nodal 
precession frequency.

\appendix

\section{Computation of ISCO energy in case of extremal Kerr-MOG BH}
In this appendix section, we would like to compute the ISCO energy for direct orbits of 
extremal Kerr-MOG BH. To do this first we should review the geodesic structure of 
time-like particle. After substituting the value of $\epsilon=-1$, one obtains the 
radial equation for time-like particle 
$$
\left(\frac{dr}{d\tau}\right)^{2} = {\cal E}^2\left(1+\frac{a^2}{r^2}+\frac{2{\cal M}a^2}{r^3}
-\frac{\alpha}{1+\alpha} \frac{a^2{\cal M}^2}{r^4}\right)
-\frac{\ell^2}{r^2}\left(1-\frac{2{\cal M}}{r}+\frac{\alpha}{1+\alpha} \frac{{\cal M}^2}{r^2}\right)
$$
\begin{eqnarray}
-2a \ell {\cal E} \left(\frac{2{\cal M}}{r^3} -\frac{\alpha}{1+\alpha} \frac{{\cal M}^2}{r^4} \right) 
-\left(1-\frac{2{\cal M}}{r}+\frac{a^2}{r^2}+\frac{\alpha}{1+\alpha} \frac{{\cal M}^2}{r^2}\right) =\chi(r)
~.\label{mradial1}
\end{eqnarray}
For circular geodesics, we know that $\chi(r)=0$ and $\frac{d \chi(r)}{dr}=0$
which gives  the energy and angular momentum for direct orbit as 
\begin{eqnarray}
{\cal E} &=& \frac{z^2-\frac{2}{1+\alpha}{\cal M}^2 z + a{\cal M}^2\sqrt{z}-\frac{\alpha}{(1+\alpha)^2}{\cal M}^4}
{\left(z+\frac{\alpha}{1+\alpha}{\cal M}^2 \right)\sqrt{ z^2-\left(\frac{\alpha+3}{\alpha+1}\right)
{\cal M}^2 z + 2a{\cal M}^2\sqrt{z}-\frac{\alpha}{(1+\alpha)^2}{\cal M}^4 }} ~.\label{ape}
\end{eqnarray}
and 
\begin{eqnarray}
\ell &=&   \frac{\sqrt{z}\left[\left(z+\frac{\alpha}{1+\alpha}{\cal M}^2 \right)^2+a^2{\cal M}^2- 2a{\cal M}^2\sqrt{z}
\right]-\frac{\alpha}{1+\alpha} a {\cal M}^4} 
{\left(z+\frac{\alpha}{1+\alpha}{\cal M}^2 \right)\sqrt{z^2-\left(\frac{\alpha+3}{\alpha+1}\right)
{\cal M}^2 z + 2a{\cal M}^2\sqrt{z}-\frac{\alpha}{(1+\alpha)^2}{\cal M}^4 }} 
\end{eqnarray}
where we have set the parameter $z={\cal M}r-\frac{\alpha}{1+\alpha}{\cal M}^2$.

To derive the direct ISCO radius, one must solve the following equation
\begin{eqnarray}
\frac{d^2\chi(r)}{dr^2}=0 
\end{eqnarray}
After long algebraic calculation, one gets 
$$
 r^2(r-6{\cal M})-3a^2r+9\left(\frac{\alpha}{1+\alpha}\right){\cal M}^2r
$$
\begin{eqnarray}
+ 8a\sqrt{{\cal M}} \left(r-\frac{\alpha}{1+\alpha}{\cal M}\right)^{3/2}
+4\left(\frac{\alpha}{1+\alpha}\right){\cal M}a^2-4 \left(\frac{\alpha}{1+\alpha}\right)^2{\cal M}^3 
&=& 0 ~.\label{isco2}
\end{eqnarray}

Now to determine the direct ISCO radius of extremal Kerr-MOG BH one should substitute
$r= \frac{y}{{\cal M}}+\frac{\alpha}{1+\alpha}{\cal M}$ in the above equation then one gets 
$$
y^3-3\left(\frac{2+\alpha}{1+\alpha}\right){\cal M}^2y^2+
3{\cal M}^2 \left[\left(\frac{\alpha}{1+\alpha}\right)^2
{\cal M}^2-\left(\frac{\alpha}{1+\alpha}\right) {\cal M}^2-a^2 \right] y
$$
\begin{eqnarray}
 + 8a {\cal M}^2y^\frac{3}{2}
+\left(\frac{\alpha}{1+\alpha}\right){\cal M}^4 \left[\left(\frac{\alpha}{1+\alpha}\right)^2{\cal M}^2
-\left(\frac{\alpha}{1+\alpha}\right){\cal M}^2+a^2\right] &=& 0
\end{eqnarray}
This is basically a sixth order polynomial equation. In the extremal limit the above equation can be written as
\begin{eqnarray}
\left(\sqrt{y}-\frac{{\cal M}}{\sqrt{1+\alpha}} \right)^3
\left[\left(\sqrt{y}+\frac{{\cal M}}{\sqrt{1+\alpha}} \right)^3-3\left(\sqrt{y}+\frac{{\cal M}}{\sqrt{1+\alpha}} \right)
+2 \frac{{\cal M}}{\sqrt{1+\alpha}} \right] &=& 0
\end{eqnarray}
The first one gives the direct ISCO for extremal Kerr-MOG BH which  occurs at $r_{isco}={\cal M}$ 
when $\frac{J}{{\cal M}^2}\geq \frac{1}{\sqrt{2}}$.  After substituting 
the value of  $r_{isco}={\cal M}$ in Eq.~(\ref{ape}), one can easily obtain the value of ISCO 
energy for direct orbit (in the extremal limit) 
\begin{eqnarray}
{\cal E}_{isco} &=& \frac{1}{\sqrt{3-\alpha}} ~.\label{ap6}
\end{eqnarray}
In the limit $\alpha=0$, one gets the ISCO energy for extremal Kerr BH~\cite{bpt}.

\section*{Acknowledgement}
I am thankful to Prof. P. Majumdar of RMVU \& IACS for reading the manuscript  and 
giving me the valuable suggestions.

\end{document}